\documentstyle[10pt]{article}
\evensidemargin \oddsidemargin
\def\br{\begin{eqnarray}}
\def\er{\end{eqnarray}}
\def\be{\begin{equation}}
\def\ee{\end{equation}}

\def\({\left(}
\def\){\right)}

\relax

\def\O{\Omega}

\def\rlx{\relax\leavevmode}
\def\inbar{\vrule height1.5ex width.4pt depth0pt}
\def\IZ{\rlx\hbox{\sf Z\kern-.4em Z}}
\def\IR{\rlx\hbox{\rm I\kern-.18em R}}
\def\IC{\rlx\hbox{\,$\inbar\kern-.3em{\rm C}$}}
\def\one{\hbox{{1}\kern-.25em\hbox{l}}}

\begin{document}
\begin{flushright}
IFT--P.065/98\\ 
\end{flushright}
\begin{center}
{\Large \bf  Classical and Quantum $V$-algebras}
\\
\end{center}
\vspace{1.0cm}
{\bf J. F. Gomes, F. E. Mendon\c{c}a da Silveira, A. H. Zimerman}
{\it Instituto de F\'{\i}sica Te\'orica - UNESP,
 Rua Pamplona, 145 , 01405-900, Sao Paulo - SP, Brazil}\\
 jfg@axp.ift.unesp.br, eugenio@axp.ift.unesp.br, zimerman@axp.ift.unesp.br

\vspace{0.5cm}

\noindent
{\bf G. M. Sotkov}\footnote{On leave of absence from the Institute for Nuclear
 Research and Nuclear Energy, Bulgarian Academy of Sciences, 1784, Sofia}
{\it Departamento de Campos e Part\'{\i}culas - CBPF
 Rua Dr. Xavier Sigaud, 150 -Urca
 22290-180, Rio de Janeiro - RJ, Brazil}\\
  sotkov@cbpfsu1.cat.cbpf.br

\vspace{1.0cm} 
     
$\frac{}{}$

\vspace{0.7cm}

\noindent
{\bf ABSTRACT}
The problem of the classification of the extensions of the Virasoro algebra is
discussed. It is shown that all $H$-reduced $\hat{\cal G}_{r}$-current algebras
belong to one of the following basic algebraic structures: local quadratic
$W$-algebras, rational $U$-algebras, nonlocal $V$-algebras, nonlocal quadratic
$WV$-algebras and rational nonlocal $UV$-algebras. The main new features of the
quantum $V$-algebras and their heighest weight representations are demonstrated
on the example of the quantum $V_{3}^{(1,1)}$-algebra.

\vspace{0.7cm}

\section{Introduction}


The concept of algebras and groups of symmetries (finite and infinite, Lie and
non-Lie etc) is, by no means, the key stone of all the field and string theories
of unification of the interactions. An impressive example of the computational
power of the algebraic methods, however, is provided by the theory of the
{\it second order phase transitions}, in two dimensions. It turns out \cite{BPZ} that the
complete nonperturbative description of the critical behaviour of a class of
2-$D$ statistical mechanics models is given by the highest weight ({\it h.w.})
unitary representations $\{ c(m),\Delta_{p,q}(m)\}$ of the Virasoro algebra
\begin{eqnarray}
[L_{n},L_{l}]=(n-l)L_{n+l}+\frac{c}{12}n(n^{2}-1)\delta_{n+l,0},
& & 
n,l=0,\pm 1,\pm 2,...
\label{1}
\end{eqnarray}
In words, all the physical data of the critical model - the exact values of the
critical exponents, the partition function, all the correlation functions etc -
are encoded in the representation theory of the algebra (\ref{1}). The exact
formulation of the above statement is as follows:

\vspace{0.3cm}

$\bullet$ {\bf Physical data} (critical $RSOS$ models on 2-$D$
planar lattice \cite{ABF}): For each fixed $m=3,4,5,...$, the
$m-th$ $RSOS$ model is defined by attaching to each site $\vec{i}$ a
height $l_{i}$ of length $l_{i}=1,2,...,m$ under the condition that the maximal length
difference of the nearest neighbours ({\it n.n.}) heights is one, i. e.,
$|l_{\vec{i}}-l_{\vec{i+1}}|=1$. The only {\it n.n.}$^{\prime}$s interact and
the energy of a given configuration is
\begin{eqnarray}
H=-\sum_{<ij>}J_{ij}l_{i}l_{j}+h\sum_{i}l_{i}.
\nonumber
\end{eqnarray}

The partition function
${\cal Z}(T,h)=Tr_{\cal H}\exp \left( -\frac{H}{kT}\right)$ ($Tr_{\cal H}$
denotes a sum over all allowed height configurations), found in \cite{ABF} shows
that, at a certain critical temperature $T=T_{c}(m)$, the $m-th$
$RSOS$ exhibits a second order phase transition. This means that, at
$\tau =\frac{T-T_{c}}{T_{c}}\rightarrow 0$, all thermodynamical characteristics
of the model have {\it power-like singularities}:

\begin{center}
$C_{V}\sim \tau^{-\alpha}$, 
\ \
$M\sim \tau^{-\beta}$,
\ \
$X\sim \tau^{-\gamma}$, ...
\end{center}

\noindent The critical exponents $\alpha (m)$, $\beta (m)$, $\gamma (m)$, ...,
turns out to be certain nonnegative rational numbers \cite{ABF}. For example,
the $m=3$ odd lattice ($l_{i}=1,3$) model is equivalent to the Ising model and
it has $\alpha =0$, $\beta =\frac{1}{8}$; the $m=5$ ($l_{i}=1,3,5$) describes
the 3-state Potts model etc.

\vspace{0.3cm}

$\bullet$ {\bf Mathematical data} (\cite{Kac1}): For each fixed $c$, the
{\it h. w.} states $|\Delta ,c>$ of the Virasoro algebra (\ref{1}) are defined
by requiring
\begin{eqnarray}
L_{0}|\Delta ,c>=\Delta |\Delta ,c>,
\ \
L_{n}|\Delta ,c>=0,
\ \
n>0.
\label{2}
\end{eqnarray}
The {\it h. w.} unitary representations\footnote{The unitary
condition was found in ref \cite{FQS1}.} of (\ref{1}) are given by
\begin{eqnarray}
\Delta_{p,q}(m)=\frac{[(m+1)p-mq]^{2}-1}{4m(m+1)},
& &
c(m)=1-\frac{6}{m(m+1)},
\label{3}
\end{eqnarray}
where $1\leq p\leq m-1$, $1\leq q\leq p$, $m=3,4,...$.

\vspace{0.3cm}

$\bullet$ {\bf Identification} (\cite{BPZ}, \cite{Dots}, \cite{Huse}): The
{\it scale invariance} of 2-$D$ statistical models, at the critical point
$T=T_{c}$, is shown to be a part of a larger group of conformal transformations
$(z,\bar{z})\rightarrow (f(z),\bar{f}(\bar{z}))$, which governes the critical
behaviour of these models in the continuous (thermodynamical) limit. Therefore,
the critical $RSOS$ models can be described in terms of certain {\it conformal
invariant 2-$D$ field theories} ($CFT^{\prime}$s)
($l_{\vec{i}}\equiv l_{i_{1}i_{2}}^{(m)}\rightarrow l^{(m)}(z,\bar{z}))$. The
symmetries of these $CFT^{\prime}$s are generated by the two components $T(z)$
and $\bar{T}(\bar{z})$ of the conserved traceless stress-tensor $T_{\mu \nu}$.
Its short distance operator-product expansion ($OPE$) is completely determined
by the symmetry
\begin{eqnarray}
T(z_{1})T(z_{2})=\frac{c/2}{z_{12}^{4}}+\frac{2T(z_{2})}{z_{12}^{2}}
+\frac{\partial_{2}T(z_{2})}{z_{12}}+O(1)
\label{4}
\end{eqnarray}
and the same for $\bar{T}(\bar{z})$. Introducing the corresponding conserved
charges $L_{n}=\oint T(z)z^{n+1}dz$, where $n=0,\pm 1,\pm 2,...$ (and the charge
$\bar{L}_{n}$ for $\bar{T}(\bar{z})$), and substituting them in (\ref{4}), we
realize that the algebra of the charges $L_{n}^{\prime}$s (and
$\bar{L}_{n}^{\prime}$s) contains two (mutualy commuting) {\it Virasoro
algebras} (\ref{1}). As a consequence, the (Hilbert) space of states of such
quantum $CFT$ can be constructed as a tensor product of two {\it h. w.}
representation spaces (\ref{2}), (with\, \, $c(m)=\bar{c}(m)$):\, \, \, \,
$|\Delta ,\bar{\Delta},c>=P(|\Delta ,c>\otimes |\bar{\Delta},c>)$, where $P$ is
denoting an appropriate projection on the subspace of states in
${\cal H}_{\Delta ,c}\otimes {\cal H}_{\bar{\Delta},c}$, satisfying certain
physical conditions - crossing symmetry, semi-locality etc - see refs \cite{BPZ}
and \cite{Dots}. To each {\it h. w.} state, one can make, in correspondence, a
{\it primary} field $\phi_{\Delta ,\bar{\Delta}}(z,\bar{z})$ of {\it spin}
$s=\Delta -\bar{\Delta}$ and {\it dimension} $d=\Delta +\bar{\Delta}$ such that
$|\Delta ,\bar{\Delta},c>=\phi_{\Delta ,\bar{\Delta}}(0,0)|0>$. One of the most
important properties of the primary fields
$\phi_{\Delta_{p,q}}\equiv \phi_{p,q}$, from the Kac-table (\ref{3}), is that,
together with the standard conformal Ward identities
\begin{eqnarray}
T(z_{1})\phi_{p,q}(z_{2})|0>
=\left( \frac{\Delta_{p,q}}{z_{12}^{2}}\phi_{p,q}(z_{2})
+\frac{1}{z_{12}}\partial_{2}\phi_{p,q}(z_{2})+O(1)\right) |0>,
\label{5}
\end{eqnarray}
it has to satisfy the so-called {\it null vector conditions}, which say for
$p=2$, $q=1$ appears to be in the form
\begin{eqnarray}
\left\{ L_{-1}^{2}-\frac{2}{3}(1+\Delta_{21}(m)L_{-2}\right\} |\Delta_{21},c>=0.
\label{6}
\end{eqnarray}
Eqns. (\ref{5}) and (\ref{6}) allow us to calculate the structure constants of
the $OPE^{\prime}$s $\phi_{p_{1}q_{1}}(z_{1})\phi_{p_{2}q_{2}}(z_{2})$, as well
as the exact 4-point (and $n$-point) correlation functions
$<\phi_{p_{1}q_{1}}(1)\phi_{p_{2}q_{2}}(2)\cdot ...\cdot \phi_{p_{n}q_{n}}(n)>$.
Finally, the identification with the $RSOS$ models is done by comparing the
$ABF$-exponents $\alpha ,\beta ,\gamma ,...$\cite{ABF}, with the Kac-dimensions
(\ref{3}). For the Ising model ($m=3$,\, \, $c(m)=\frac{1}{2}$), we have
$\alpha =0$,\, \, $\beta =\frac{1}{8}$ \, \, and \, \, 
$\Delta_{21}(3)=\frac{1-\alpha}{2-\alpha}=\frac{1}{2}$,\, \,
$\Delta_{22}(3)=\frac{\beta}{2-\alpha}=\frac{1}{16}$.

\vspace{0.3cm}

$\bullet$ {\bf The problem of classification of 2-$D$ universality classes}. The
purely algebraic description of the critical $RSOS$ models address the question
whether one can find appropriate infinite algebras, which representation
theories provide the exact solutions for all known 2-$D$ critical statistical
models having second order phase transition. The algebras, we are looking for,
have to contain the Virasoro algebra (\ref{1}) as a subalgebra. The following
three examples of extended Virasoro algebras are to ilustrate the main features
of the {\it new algebraic structures} one need to introduce in the description
of the universality classes, in two dimensions.

\vspace{0.2cm}

{\bf Example 1.1}. {\bf $N=1$ supersymmetric Virasoro algebra}\cite{NS}: An
infinite Lie super-algebra, containing together with the bosonic Virasoro
generators $L_{n}$, an infinite set of fermionic ones $G_{s}$
($s\in \frac{1}{2}Z$)
\begin{eqnarray}
[L_{n},G_{s}]=\left( \frac{n}{2}-s\right) G_{n+s},& &
\left[ G_{s},G_{t}\right]_{+}
=2L_{s+t}+\frac{c}{2}\left( s^{2}-\frac{1}{4}\right) \delta_{s+t,0},
\label{7}
\end{eqnarray}
where $[\frac{}{},\frac{}{}]_{+}$ denotes the {\it anticommutator} of
$G_{t}^{\prime}$s. The {\it h. w.} representations of (\ref{7})\cite{Kac1}
\begin{eqnarray}
c(m)=\frac{3}{2}\left( 1-\frac{8}{m(m+2)}\right) ,& &
\Delta_{pq}(m)=\frac{[(m+2)p-mq]^{2}-2}{12m(m+2)},
\label{8}
\end{eqnarray}
where $m=4,5,...$ and $1\leq p\leq m-2$, $1\leq q\leq p$, give rise to a family
of superconformal minimal models \cite{FQS2}, \cite{KBT} which describe the
critical behaviour of $k=2$ generalized $RSOS$ models. The difference with the
standard ($k=1$) $RSOS$ is that now the allowed {\it maximal} length difference
between {\it n. n.} heights $l_{\vec{i}}$ and $l_{\vec{i+1}}$ is $k=2$. As it is
evident from (\ref{7}), the $k=2$ critical $RSOS$ possess symmetry larger than
the conformal one. The stress-tensor ($T,\bar{T}$) and the new spin $\frac{3}{2}$
supercurrent ($G,\bar{G}$), $G_{n}=\oint z^{n+\frac{1}{2}}G(z)dz$, generate
2-$D$ superconformal transformations.

The critical $k-RSOS$ models for $k=3,4,...$ require {\it fractional} spin
$\frac{l}{k}$ extensions of the Virasoro algebra \cite{KMQ, Rav}, \cite{ZF1}.
Our next example represents the main features of such ``parafermionic type''
algebras.

\vspace{0.2cm}

{\bf Example 1.2}. {\bf $Z_{N}$ Parafermionic algebra} \cite{ZF1}, \cite{ZF2}:
The $Z_{N}$ generalizations of the Ising ($Z_{2}$) and Potts ($Z_{3}$) models
are lattice spin models, where each site ($i$) is occupied by a
``spin variable'' $\sigma (i)$ that takes values
$\sigma_{(i)}^{l}=\exp \left( \frac{2\pi \imath l}{N}\right) $ in the discrete
group $Z_{N}$. To describe (multi) critical behaviour of these models, one has
to consider, together with $T(z)$, a set of $N-1$ new conserved currents
$\psi_{l}^{+}(z)=\psi_{N-l}^{-}(z)$, where $l=1,2,...,N-1$, of spins
$s_{l}=\frac{l(N-l)}{N}$, with $OPE^{\prime}$s in the form \cite{ZF1},
\cite{ZF2}
\begin{eqnarray}
\psi_{1}^{\pm}(z_{1})\psi_{1}^{\pm}(z_{2})
&=&c_{11}z_{12}^{-\frac{2}{N}}(\psi_{2}^{\pm}(z_{2})+O(z_{12})),
\nonumber
\\
\psi_{1}^{+}(z_{1})\psi_{1}^{-}(z_{2})&=&z_{12}^{\frac{2}{N}}
\left( \frac{1}{z_{12}^{2}}+\frac{N+2}{N}T(z_{2})+O(z_{12})\right) ,
\label{9}
\end{eqnarray}
where $c_{11}=\sqrt{\frac{2(N-1)}{N}}$.Introducing the parafermionic ($PF$)
conserved charges in (\ref{9})
\begin{eqnarray}
A_{\frac{1\mp l}{N}+n}^{\pm}\phi_{l}(0)
&=&\oint dz\psi_{1}^{\pm}(z)z^{\mp \frac{l}{N}+n}\phi_{l}(0),
\nonumber
\\
\psi_{1}^{\pm}(z\exp (2\pi \imath ))\phi_{l}(0)
&=&\exp (\frac{2\pi \imath l}{N})\psi_{1}^{\pm}(z)\phi_{l}(0),
\nonumber
\end{eqnarray}
we derive the $Z_{N}$ $PF$-extension of the Virasoro algebra (\ref{2})
\begin{eqnarray}
&\frac{}{}&\sum_{p=0}^{\infty}C_{\left( \frac{2}{N}\right) }^{p}
\left( A_{\frac{3\mp l}{N}-p+m}^{\pm}A_{\frac{1\mp l}{N}+p+n}^{\pm}
-A_{\frac{3\mp l}{N}-p+n}^{\pm}A_{\frac{1\mp l}{N}+p+m}^{\pm}\right) =0,
\nonumber
\\
&\frac{}{}&\sum_{p=0}^{\infty}C_{\left( -\frac{2}{N}\right) }^{p}
\left( A_{m-\frac{1+l}{N}-p}^{+}A_{\frac{1+l}{N}+n+p-1}^{-}
+A_{n-\frac{1-l}{N}-p-1}^{-}A_{\frac{1-l}{N}+p+m}^{+}\right)
\nonumber
\\
&\frac{}{}&=\frac{N+2}{N}L_{m+n-1}+\frac{1}{2}\left( n-1+\frac{l}{N}\right)
\left( \frac{l}{N}+m-2\right) \delta_{m+n-1,0}
\label{10}
\end{eqnarray}
of central charge $c(N)=2\frac{N-1}{N+2}$, where $N=2,3,...$, and structure
constants $C_{(r)}^{p}=\frac{\Gamma (p-r)}{p!\Gamma (-r)}$. The {\it h. w.}
representations of this infinite associative algebra, found in ref \cite{ZF1},
are of dimensions $\Delta_{l}=\frac{l(N-l)}{2N(N+2)}$ and $Z_{N}$ charge - 
$l=1,2,...,N-1$ for the ({\it order parameter}) fields $\sigma_{l}(z,\bar{z})$;
and $\Delta_{j}=\frac{j(j+1)}{N+2}$ where
$j=1,2,...\leq \left[ \frac{N}{2}\right] $ are the dimensions for the $Z_{N}$
neutral ({\it energy operator}) fields $\epsilon_{j}(z,\bar{z})$. It is
important to note that the {\it origin} of the fact that in the $PF$-algebra
(\ref{10}), the Lie commutator $[a,b]=ab-ba$ is replaced by an infinite sum of
bilinears $A^{\pm}A^{\mp}$ is in the {\it branch cut singularities}
$z^{\pm \frac{2}{N}}$ ($N\geq 3$) in the $OPE^{\prime}$s (\ref{9}). These types
of singularities are a consequence of the {\it fractal spins}
$s_{1}^{\pm}=1-\frac{1}{N}$ of the $PF$-currents $\psi_{1}^{\pm}$. Observe that
for $N=2$ (and for half-integer spins, in general), the $OPE^{\prime}$s have odd
poles $z_{12}^{-1}$ (or $z_{12}^{-2s-1}$) singularities, which lead to
anticommutators $[a,b]_{+}=ab+ba$. For integer spins $s=1,2,3,...$, the leading
singularities in the $OPE^{\prime}$s are even poles $z_{12}^{-2s}$
(as in eqn. (\ref{4})) and they give rise to the standard Lie commutators.

\vspace{0.2cm}

{\bf Example 1.3}. {\bf $W_{3}$-Zamolodchikov algebra} \cite{Zam}, \cite{ZF3}:
The most important property of the spin 3 extension of the Virasoro algebra
(generated by $T(z)$ and $W(z)$ of spin $s_{W}=3$) is that the commutator of the
charges $W_{n}$ of the spin 3-current $W_{n}=\oint W(z)z^{n+2}dz$ is quadratic
in the Virasoro generators $L_{n}^{\prime}$s
\begin{eqnarray}
[W_{n},W_{l}]=(n-l)[d(n,l)L_{n+l}+b\Lambda_{n+l}]
+\frac{c}{360}n(n^{2}-4)(n^{2}-1)\delta_{n+l,0},
\label{11}
\end{eqnarray}
where
\begin{eqnarray}
\Lambda_{n}
=\sum_{k=-\infty}^{\infty}:L_{k}L_{n-k}:+\frac{1}{5}f_{n}L_{n},
\ \
f_{2s}=1-s^{2},
\ \
f_{2s+1}=(1-s)(2+s)
\nonumber
\end{eqnarray}
and
\begin{eqnarray}
d(n,l)=\frac{1}{6}\left[ \frac{2}{5}(n+l+2)(n+l+3)-(n+2)(l+2)\right] ,
& & 
b=\frac{16}{22+5c}.
\nonumber
\end{eqnarray}
The {\it h. w.} states $|\Delta ,w,c>$ of this {\it non-Lie associative} algebra
are defined by
\begin{eqnarray}
L_{0}|\Delta ,w,c>=\Delta |\Delta ,w,c>,
& &
W_{0}|\Delta ,w,c>=w|\Delta ,w,c>,
\nonumber
\er
\begin{center}
$L_{n}|\Delta ,w,c>=W_{n}|\Delta ,w,c>=0$,\quad 
$n>0$.
\end{center}
Its {\it h. w.} unitary representations
$c(m)=2\left( 1-\frac{12}{m(m+1)}\right)$, where $m=4,5,...$, and
$\Delta_{p_{i}q_{i}}(m)$, $w_{p_{i}q_{i}}(m)$ ($i=1,2$) found in ref \cite{ZF3}
give rise to a family of $Z_{3}$ symmetric $CFT^{\prime}$s that provide the
exact solutions for a new class of critical statistical models. The simplest
representative $m=4$ of this class is again the critical 3-states Potts model.

The above examples of three different associative extensions of the Virasoro
algebra, (\ref{7}), (\ref{10}), (\ref{11}), suggest the following
{\it organization} of the list of all known infinite algebras (and their 2-$D$
$CFT^{\prime}$s):

\vspace{0.3cm}

{\bf (i)} {\it Lie-algebraic extensions}: Conformal current (affine)
$\hat{\cal G}_{r}$-algebras \cite{Kac-Moody}:
\begin{eqnarray}
[J_{n}^{a},J_{l}^{b}]=\imath f^{abc}J_{n+l}^{c}+kn\delta^{ab}\delta_{n+l,0},
\label{12}
\end{eqnarray}
where $n,l=0,\pm 1,\pm 2,...$; $a,b=1,2,...,dim$ ${\cal G}$; $f^{abc}$ are the
structure constants of an arbitrary (finite dimensional) semisimple Lie algebra
${\cal G}_{r}$; $k$ is called the level of $\hat{\cal G}_{r}$. Its generators
are the conserved charges of the spin $s=1$ chiral current
$J^{a}(z)=\sum_{n=-\infty}^{\infty}z^{-n-1}J^{a}_{n}$, which also satisfy
$[L_{n},J_{l}^{a}]=-lJ_{n+l}^{a}$. The {\it h. w.} representations of (\ref{12}) (and its
$CFT^{\prime}$s ) were constructed in refs \cite{KZ, GW}.

\vspace{0.2cm}

{\bf (ii)} {\it Lie-superalgebraic extensions}: The  $N=1$ superVirasoro
algebra (\ref{7}); $N=2,3,4$ superconformal algebras \cite{Ad, BFK, DiVec1};
the affine $\hat{\cal G}_{r}$-superalgebras, where ${\cal G}_{r}$ is an
arbitrary rank $r$ finite dimensional superalgebra; $N=1$ superconformal current
$\hat{\cal G}_{r}$-algebras \cite{KT, DiVec2}, with generators $J_{n}^{a}$ and
$\psi_{n}^{a}$ determined by (\ref{7}), (\ref{12}) and
$[J_{n}^{a},\psi_{l}^{b}]=f^{abc}\psi_{n+l}^{c}$,
$[\psi_{n}^{a},\psi_{l}^{b}]_{+}=k\delta^{ab}\delta_{n+l,0}$.

\vspace{0.2cm}

{\bf (iii)} {\it $PF$-extensions}: The $Z_{N}$ (and $D_{2N}$)-$PF$ algebra
(\ref{10}) and its $(p,M)$-general\-izations \cite{ZF1}, by considering $PF$
currents of spins $s_{l}=p\frac{l(N-l)}{N}+M_{l}$; Gepner$^{\prime}$s
${\cal G}_{r}$-parafermions \cite{Gepn}.

\vspace{0.2cm}

{\bf (iv)} {\it Quadratic $W$-algebras}: The $W_{n}$-algebras \cite{FL},
\cite{BBSS}, \cite{BG} generated by the charges of the spin
$s=2,3,...,n$-currents, and the more general $W{\cal G}_{n}$ \cite{FL}; the
$W_{n}^{(l)}$-algebras \cite{Bersh}, \cite{Poly1}, \cite{O'Rai2}; the
supersymmetric $W_{n}$-algebras etc.

\vspace{0.3cm}

To complete our table of extended Virasoro algebras, we have to add the family
of the recently discovered classical Poisson brackets {\it nonlocal and
nonlinear (quadratic)} $V$-algebras \cite{Bil1}, \cite{Bil2}, \cite{GSZ1},
\cite{GSZ2}.

{\bf (v)} {\it $V$-algebras}: The simplest example is given by
$V_{3}^{(1,1)}\equiv VA_{2}^{(1,1)}$-algebra \cite{GSZ1}, \cite{GSZ2}, generated
by one local spin 2 $T(\sigma )$ (the stress-tensor) and two spin
$\frac{3}{2}$-non local currents $V^{\pm}(\sigma )$:
\begin{eqnarray}
\{ T(\sigma ),V^{\pm}(\sigma^{\prime})\}
=\frac{3}{2}V^{\pm}(\sigma^{\prime})\partial_{\sigma^{\prime}}
\delta (\sigma -\sigma^{\prime})
+\partial_{\sigma^{\prime}}V^{\pm}(\sigma^{\prime})
\delta (\sigma -\sigma^{\prime}),
\nonumber
\end{eqnarray}
\begin{eqnarray}
\{ V^{\pm}(\sigma ),V^{\mp}(\sigma^{\prime})\}
=\pm \frac{2}{k}\delta^{\prime \prime}(\sigma -\sigma^{\prime})
\mp \frac{2}{k}T(\sigma^{\prime})\delta (\sigma -\sigma^{\prime})
+\frac{3}{2k^{2}}V^{\pm}(\sigma )V^{\mp}(\sigma^{\prime})
\epsilon (\sigma -\sigma^{\prime}),
\nonumber
\end{eqnarray}
\begin{eqnarray}
\{ V^{\pm}(\sigma ),V^{\pm}(\sigma^{\prime})\}
=-\frac{3}{2k^{2}}V^{\pm}(\sigma )V^{\pm}(\sigma^{\prime})
\epsilon (\sigma -\sigma^{\prime}).
\label{13}
\end{eqnarray}
where $\epsilon (\sigma )=sign$ $\sigma$. The $V_{3}^{(1,1)}$ is the first
member of the $VA_{n}^{(1,1)}$-family of $V$-algebras, spanned by two non-local
currents $V_{(n)}^{\pm}$ of spins $s=\frac{n+1}{2}$ and $n-1$ local currents
$W_{n-l+2}$ of spins $s_{l}=n-l+2$, where $l=1,2,...,n$. The Bilal$^{\prime}$s
$VB_{2}$-algebra \cite{Bil1} is quite similar to (\ref{13}), but
$V^{\pm}$-currents have spin $s^{\pm}=2$ in this case.

\vspace{0.3cm}

Our main purpose, in what follows, is the construction of the {\it quantum}
$V_{n+1}$-algebras and their minimal conformal models (i. e., their {\it h. w.}
representations). The most important result is that the classical spins
$s^{cl}=\frac{n+1}{2}$, of the nonlocal currents $V^{\pm}$, gets renormalized,
i. e., $s^{qu}=\frac{n+1}{2}\left( 1-\frac{1}{2k+n+1}\right)$ and their algebra
shares the main patterns of the $PF$-algebras. While the quantum local currents
$W_{n-l+2}$ manifest properties similar to the $W_{n}$-al\-ge\-bras. Therefore,
the quantum $V_{n+1}^{(1,1)}$-algebras represent an appropriate
{\it unification} of the features of the $Z_{2k+3}$ $PF$-algebra with the
$W_{n+1}$-one.


\section{Constrained ${\cal G}_{r}$-current algebras}


The list of the five known families of extended Virasoro algebras we have made,
however, does {\it not} solve the problem of the {\it classification of 2-$D$
universality classes} (i. e., all allowed critical behaviours in two dimensions). We need a
method of exhausting all the possible extensions of the Virasoro algebra. The
hint is coming from the fact that all the considered algebras\footnote{The
supersymmetric extensions arise from the constrained superconformal current
algebras} - the Virasoro-one, the $PF$-, the $W_{n}$- and $V_{n}$-ones - can be
obtained by imposing a specific set of constraints on the currents of certain
$\hat{\cal G}_{r}$-current algebras ($SL(2,R)$, for (\ref{1}) and (\ref{10}),
and $SL(3,R)$, for (\ref{11}) and (\ref{13}) etc):
\begin{eqnarray}
\{ J^{a}(\sigma ),J^{b}(\sigma^{\prime})\}
=\imath f^{abc}J^{c}(\sigma^{\prime})\delta (\sigma -\sigma^{\prime})
+k\partial_{\sigma^{\prime}}\delta (\sigma -\sigma^{\prime}).
\label{14}
\end{eqnarray}
It suggests that the desired classification of the extended Virasoro algebras
can be reached by the methods of the Hamiltonian reduction \cite{Poly1},
\cite{BO}, \cite{O'Rai1}, i. e., by considering all  consistent
sets of constraints on the currents $J^{a}(\sigma )\in \hat{\cal G}_{r}$
\begin{eqnarray}
\hat{J}(\sigma )=g^{-1}\partial g
=\sum_{\rm all roots}J_{\{ \alpha \} }E_{\{ \alpha \} }
+\sum_{i=1}^{r}J_{i}\frac{\vec{\alpha}_{i}\cdot \vec{H}}{\alpha_{i}^{2}}
\nonumber
\end{eqnarray}
where $E_{\{ \alpha \} }$,
$h_{i}=\frac{\vec{\alpha}_{i}\cdot \vec{H}}{\alpha_{i}^{2}}$ are the generators
of the finite Lie algebra ${\cal G}_{r}$. Therefore, the question now is 
{\it whether and how} one can {\it classify} all constraints to be imposed on
$\hat{J}(\sigma )$.

We start with few selected examples of constrained $SL(n,R)$ ($n=2,3,4$)
algebras, which demonstrate the way the {\it algebraic structure} of the
reduced algebras depends on the specific choice of the constraints.


{\bf Example 2.1}. {\bf $SL(2,R)$ reductions}.


{\bf (1a)} {\it $A_{1}/{\cal N}_{+}\equiv$ Virasoro algebra}: Take
$J_{\alpha}=1$ as a constraint and $J_{1}(\equiv J_{0})=0$ as its gauge fixing
(i. e., $J_{1}$ is the canonically conjugated momentum of $J_{\alpha}$, since
$\{ J_{\alpha}(\sigma ),J_{1}(\sigma^{\prime})\}
=-J_{\alpha}(\sigma )\delta (\sigma -\sigma^{\prime})
\approx \delta (\sigma -\sigma^{\prime})$).\footnote{An equivalent explanation
of the $J_{1}=0$ condition (which is not a constraint) is that, due to the
residual gauge transformation $h=\exp (\beta (z))E_{-\alpha}$,
$J^{\prime}=h^{-1}Jh+kh^{-1}\partial h$, which leaves invariant the constraint
$J_{\alpha}=1$, one can make $J_{1}^{\prime}=0$, by an appropriate choice of
$\beta (z)$.} Under these conditions, the classical Poisson bracket ($PB$)
algebra of the remaining current $J_{-\alpha}\equiv T$ can be derived from eqn.
(\ref{11}), by calculating the corresponding Dirac brackets
\begin{eqnarray}
\{ T(\sigma ),T(\sigma^{\prime})\}_{D}
=\frac{k^{2}}{2}\delta^{\prime \prime \prime}(\sigma -\sigma^{\prime})
-2T(\sigma^{\prime})\delta^{\prime}(\sigma -\sigma^{\prime})
+\partial_{\sigma^{\prime}}T(\sigma^{\prime})
\delta (\sigma -\sigma^{\prime}),
\label{15}
\end{eqnarray}
which is nothing, but the {\it classical} $PB^{\prime}$s Virasoro algebra.
Another form of the Dirac method, proposed by Polyakov \cite{Poly1} consists in
imposing the constraints and gauge fixing conditions on the infinitesimal
$\hat{\cal G}_{r}$-gauge transformations
\begin{eqnarray}
\delta_{\epsilon}J^{a}(\sigma )=f^{abc}J^{c}(\sigma )\epsilon^{b}(\sigma )
+\frac{k}{2}\partial_{\sigma}\epsilon^{a}.
\label{16}
\end{eqnarray}
Next, we solve the $\delta_{\epsilon}J_{\alpha}=\delta J_{1}=0$ equations for
the redundant gauge parameters $\epsilon_{1}$ and $\epsilon_{\alpha}$
\begin{eqnarray}
\epsilon_{1}=\frac{k}{2}\partial \epsilon ,
\ \
\epsilon_{\alpha}=-\frac{k^{2}}{2}\partial^{2}\epsilon +J_{-\alpha}\epsilon ,
\ \
\epsilon \equiv \epsilon_{-\alpha},
\nonumber
\end{eqnarray}
and substituting them in $\delta J_{-\alpha}$, we find
\begin{eqnarray}
\delta_{\epsilon}J_{-\alpha}=-\frac{k^{2}}{2}\partial^{3}\epsilon
+2J_{-\alpha}\partial \epsilon +\partial J_{-\alpha}\epsilon ,
\nonumber
\end{eqnarray}
i. e., the functional form of eqn. (\ref{15}).

\vspace{0.2cm}

{\bf (1b)} {\it $A_{1}/U(1)\equiv$ classical $PF$-algebra}: Take
$J_{0}(\equiv J_{1})=0$ as a constraint (no residual gauge transformations
exists). In this case, as a consequence of eqn. (\ref{14}), we have
$\{ J_{0}(\sigma ),J_{0}(\sigma^{\prime})\}
=\partial_{\sigma^{\prime}}\delta (\sigma -\sigma^{\prime})$ ($k=2$). To find
the Dirac brackets of the $J_{\pm \alpha}^{\prime}$s, we have to invert the
$\partial_{\sigma}$-operator, i. e.,
$\partial_{\sigma}(\partial^{-1}_{\sigma^{\prime}})
=\delta (\sigma -\sigma^{\prime})$, hence
$\partial^{-1}_{\sigma^{\prime}}
=\frac{1}{2}\epsilon (\sigma -\sigma^{\prime})$) and thus to introduce
{\it nonlocal} $\epsilon (\sigma )$-terms in the
$J_{\pm \alpha}\equiv V^{\pm}$-algebra
\begin{eqnarray}
\{ V^{\pm}(\sigma ),V^{\pm}(\sigma^{\prime})\}_{D}
&=&-V^{\pm}(\sigma )V^{\pm}(\sigma^{\prime})
\epsilon (\sigma -\sigma^{\prime}),
\nonumber
\\
\{ V^{+}(\sigma ),V^{-}(\sigma^{\prime})\}_{D}
&=&\partial_{\sigma^{\prime}}\delta (\sigma -\sigma^{\prime})
+V^{+}(\sigma )V^{-}(\sigma^{\prime})
\epsilon (\sigma -\sigma^{\prime}).
\label{17}
\end{eqnarray}
Following the Polyakov method, we get
\begin{eqnarray}
\epsilon_{0}(\sigma )=-\int \epsilon (\sigma -\sigma^{\prime})
(V^{+}(\sigma^{\prime})\epsilon^{-}(\sigma^{\prime})
-V^{-}(\sigma^{\prime})\epsilon^{+}(\sigma^{\prime}))d\sigma^{\prime},
\nonumber
\end{eqnarray}
and pluging it back in the $\delta_{\epsilon^{\pm}}V^{\pm}$-transformations, we
arrive at eqn. (\ref{17}). The reason to call this {\it nonlocal} $PB$ algebra
as a {\it classical parafermionic} one is that an appropriate
$N\rightarrow \infty$ limit of the (quantum) $PF$ $OPE^{\prime}$ (\ref{9})
reproduces exactly eqn. (\ref{17}), as we will demonstrate, in detail, in the
next section.


{\bf Example 2.2}. {\bf Constrained $SL(3,R)$-algebras}.


{\bf (2a)} {\it $A_{2}/{\cal N}_{+}\equiv W_{3}$-algebra}: In this case,
${\cal N}_{+}=\{ E_{\alpha_{1}},E_{\alpha_{2}},E_{\alpha_{1}+\alpha_{2}}\}$ and
$J_{\alpha_{i}}=1$, $J_{\alpha_{1}+\alpha_{2}}=0$ are the constraints;
$J_{i}=J_{-\alpha_{1}}=0$ are the gauge fixing conditions, in Drinfeld-Sokolov
gauge. The classical $W_{3}$-algebra, generated by one spin 2
$T(z)\equiv J_{-\alpha_{2}}(z)$ and one spin 3
$W_{3}(z)=J_{-\alpha_{1}-\alpha_{2}}-\frac{1}{2}\partial J_{-\alpha_{2}}$
currents, has the form ($k=2$)
\begin{eqnarray}
\{ T(\sigma ),W_{3}(\sigma^{\prime})\}
=3W_{3}(\sigma^{\prime})\partial_{\sigma^{\prime}}
\delta (\sigma -\sigma^{\prime})
+2\partial_{\sigma^{\prime}}W_{3}(\sigma^{\prime})
\delta (\sigma -\sigma^{\prime}),
\nonumber
\end{eqnarray}
\begin{eqnarray}
\{W_{3}(\sigma ),W_{3}(\sigma^{\prime})\}
=-4\delta^{(v)}(\sigma -\sigma^{\prime})
+5T(\sigma^{\prime})\delta^{\prime \prime \prime}(\sigma -\sigma^{\prime})
-\frac{15}{2}\partial_{\sigma^{\prime}}T(\sigma^{\prime})
\delta^{\prime \prime}(\sigma -\sigma^{\prime})&\frac{}{}&
\nonumber
\\
-\left( T^{2}(\sigma^{\prime})
-\frac{9}{2}\partial_{\sigma^{\prime}}^{2}
T(\sigma^{\prime})\right) \delta^{\prime}(\sigma -\sigma^{\prime})
+\partial_{\sigma^{\prime}}\left( \frac{1}{2}T^{2}(\sigma^{\prime})
-\partial_{\sigma^{\prime}}^{2}T(\sigma^{\prime})\right)
\delta (\sigma -\sigma^{\prime}).&\frac{}{}&
\label{18}
\end{eqnarray}

\vspace{0.2cm}

{\bf (2b)}
{\it $A_{2}/{\cal N}_{+}^{1}\otimes U(1)\equiv V_{3}^{(1,1)}$-algebra}:
Take ${\cal N}_{+}^{(1)}=\{ E_{\alpha_{2}},E_{\alpha_{1}+\alpha_{2}}\}$,
$U(1)=\{ \vec{\lambda}_{1}\cdot \vec{H}\}$; $J_{\alpha_{2}}=1$,
$J_{\alpha_{1}+\alpha_{2}}=0$, $\sum_{i=1}^{2}\lambda_{1}^{(i)}J_{i}=0$ are the
constraints and $J_{-\alpha_{1}}=\sum_{i=1}^{2}\alpha_{2}^{i}J_{i}=0$ are the
gauge fixing conditions. The remaining currents $V^{+}=J_{\alpha_{1}}$,
$V^{-}=J_{-\alpha_{1}-\alpha_{2}}$, of spin $\frac{3}{2}$ (nonlocal), and
$T=J_{-\alpha_{2}}$, of spin 2 (local), generate the following nonlocal
$V_{3}^{(1,1)}$-algebra ($k=2$) (\ref{13}). We have to mention that if one
{\it relaxes} the $U(1)$-constraint $J=\vec{\lambda}_{1}\cdot \vec{J}=0$, then
the {\it local} currents $V^{\pm}$ ($s=\frac{3}{2}$), $T$ ($s=2$) and $J$
($s=1$) span the well known local quadratic (in $J$) Bershadsky-Polyakov
$A_{2}/{\cal N}_{+}^{(1)}\equiv$ $W_{3}^{(2)}$-algebra \cite{Bersh}.

\vspace{0.2cm}

{\bf (2c)}
{\it $A_{2}/{\cal N}_{+}^{(2)}\otimes U(1)\otimes U(1)\equiv V_{3}$
-algebra}: In this case, ${\cal N}_{+}^{(2)}=\{ E_{\alpha_{1}+\alpha_{2}}\}$,
the constraints are $J_{i}=0$, $J_{\alpha_{1}+\alpha_{2}}=0$ and the gauge
fixing ({\it g. f.}) conditions are $J_{-\alpha_{1}-\alpha_{2}}=0$, $J_{i}=0$.
The {\it nonlocal} $V_{3}$-algebra, of the four spin 1 currents
$V^{\pm}_{i}=J_{\pm \alpha_{i}}$ ($i=1,2$), has the form \cite{GSSZ}
\begin{eqnarray}
\{ V^{\pm}_{i}(\sigma ),V^{\pm}_{j}(\sigma^{\prime})\}
&=&\frac{1}{2k^{2}}[V^{\pm}_{i}(\sigma )V^{\pm}_{j}(\sigma^{\prime})
+V^{\pm}_{i}(\sigma^{\prime})V^{\pm}_{j}(\sigma )]
\epsilon (\sigma -\sigma^{\prime}),
\nonumber
\end{eqnarray}
\begin{eqnarray}
\{ V^{+}_{i}(\sigma ),V^{-}_{j}(\sigma^{\prime})\}
&=&\delta_{ij}\partial_{\sigma^{\prime}}\delta (\sigma -\sigma^{\prime})
-\frac{1}{2k^{2}}[V^{+}_{i}(\sigma )V^{-}_{j}(\sigma^{\prime})
\nonumber
\\
&+&\delta_{ij}\sum_{s=1}^{2}V^{-}_{s}(\sigma )V^{+}_{s}(\sigma^{\prime})]
\epsilon(\sigma -\sigma^{\prime}).
\label{20}
\end{eqnarray}
The stress-tensor
$T(\sigma )=\frac{1}{2}\sum_{s=1}^{2}V^{+}_{s}(\sigma )V^{-}_{s}(\sigma )$
satisfies the standard Virasoro algebra (\ref{15}), but without a central term.

\vspace{0.2cm}

{\bf (2d)} {\it $A_{2}/{\cal N}_{+}^{(2)}\otimes U(1)\equiv V_{3}^{(2)}$
-algebra}: ${\cal N}_{+}^{(2)}$ is the same as in (2c),
$U(1)=(\lambda_{1}-\lambda_{2})^{i}J_{i}$; the constraints are
$J_{\alpha_{1}+\alpha_{2}}=1$,
$\sum_{i=1}^{2}(\lambda_{1}-\lambda_{2})^{i}J_{i}=0$ and the {\it g. f.}
conditions are $(\alpha_{1}+\alpha_{2})^{i}J_{i}=0$. The $V_{3}^{(2)}$-algebra,
of the local spin 2 stress-tensor
$T=J_{-\alpha_{1}-\alpha_{2}}
-\frac{1}{2}(J_{\alpha_{1}}J_{-\alpha_{1}}+J_{\alpha_{2}}J_{-\alpha_{2}})$, and
four nonlocal currents $V^{+}_{1}=J_{\alpha_{1}}$ ($s^{+}_{1}=\frac{1}{2}$),
$V^{-}_{1}=J_{-\alpha_{1}}-2\partial J_{\alpha_{2}}$ ($s^{-}_{1}=\frac{3}{2}$),
$V^{-}_{2}=J_{\alpha_{2}}$ ($s^{-}_{2}=\frac{1}{2}$) and
$V^{+}_{2}=J_{-\alpha_{2}}+2\partial J_{\alpha_{1}}$ ($s^{+}_{2}=\frac{3}{2}$),
takes the form \cite{GSSZ}
\begin{eqnarray}
\{ V^{\pm}_{i}(\sigma ),V^{\pm}_{j}(\sigma^{\prime})\}
=(i-j)[V^{\pm}_{\frac{1}{2}(i+j\mp 1)}(\sigma )]^{2}
\delta (\sigma -\sigma^{\prime})
+\frac{3}{8}V^{\pm}_{i}(\sigma )V^{\pm}_{j}(\sigma^{\prime})
\epsilon (\sigma -\sigma^{\prime}),
\nonumber
\end{eqnarray}
\begin{eqnarray}
\{ V^{-}_{i}(\sigma ),V^{+}_{i}(\sigma^{\prime})\}
=2V^{-}_{1}(\sigma )V^{+}_{2}(\sigma )\delta (\sigma -\sigma^{\prime})
-\frac{3}{8}V^{-}_{i}(\sigma )V^{+}_{i}(\sigma^{\prime})
\epsilon (\sigma -\sigma^{\prime}),
\nonumber
\end{eqnarray}
\begin{eqnarray}
\{ V^{-}_{1}(\sigma ),V^{+}_{2}(\sigma^{\prime})\}
&=&-4\partial_{\sigma^{\prime}}^{2}\delta (\sigma -\sigma^{\prime})
-\frac{3}{8}V^{-}_{1}(\sigma )V^{+}_{2}(\sigma^{\prime})
\epsilon (\sigma -\sigma^{\prime})
\nonumber
\\
&+&3[V^{+}_{1}(\sigma )V^{-}_{2}(\sigma^{\prime})
+V^{+}_{1}(\sigma^{\prime})V^{-}_{2}(\sigma )]
\partial_{\sigma^{\prime}}\delta (\sigma -\sigma^{\prime})
\nonumber
\\
&+&\{ T(\sigma )
+\frac{3}{2}[V^{+}_{1}(\sigma )V^{-}_{1}(\sigma )
+V^{+}_{2}(\sigma )V^{-}_{2}(\sigma )]\}
\delta (\sigma -\sigma^{\prime}),
\nonumber
\end{eqnarray}
\begin{eqnarray}
\{ V^{-}_{2}(\sigma ),V^{+}_{1}(\sigma^{\prime})\}
=\delta (\sigma -\sigma^{\prime})
-\frac{3}{8}V^{-}_{2}(\sigma )V^{+}_{1}(\sigma^{\prime})
\epsilon (\sigma -\sigma^{\prime}).
\label{21}
\end{eqnarray}
Thus, $V_{3}^{(2)}$ is an example of {\it nonlocal quadratic} (non-Lie) algebra.

\vspace{0.2cm}

{\bf (2e)} {\it $A_{2}/{\cal N}_{+}^{(1)}\equiv W_{3}^{(1,1)}$-algebra}:
In this case, the constraints are $J_{\alpha_{2}}=J_{\alpha_{1}+\alpha_{2}}=1$
and the {\it g. f.} conditions are
$\alpha_{2}^{i}J_{i}=(\alpha_{1}+\alpha_{2})^{i}J_{i}=0$. The algebra
$W_{3}^{(1,1)}$, of the local currents $J^{\pm}=J_{\pm \alpha_{1}}$
($s^{\pm}=1$) and $T_{2}=J_{-\alpha_{2}}$ ($s_{2}=2$),
$T_{12}=J_{-\alpha_{1}-\alpha_{2}}$ ($s_{12}=2$), appears to be a local
quadratic algebra \cite{GSSZ}, of $W$-type.


{\bf Example 2.3}. {\bf Constrained $SL(4,R)$-algebras}.


{\bf (3a)} {\it $A_{3}/{\cal N}_{+}\equiv W_{4}$-algebra}:
${\cal N}_{+}=\{ E_{[\alpha ]}:
[\alpha ]{\rm  are\, \, all\, \, positive\, \, roots}\} $,
the constraints are $J_{\alpha_{i}}=1$ ($i=1,2,3$),
$J_{\alpha_{1}+\alpha_{2}}=J_{\alpha_{2}+\alpha_{3}}=0$,
$J_{\alpha_{1}+\alpha_{2}+\alpha_{3}}=0$ and the {\it g. f.} conditions are
$J_{i}=0$, $J_{-\alpha_{1}}=0$, $J_{-\alpha_{2}}=0$,
$J_{-\alpha_{1}-\alpha_{2}}=0$. The algebra of the remaining currents
$T=J_{-\alpha_{3}}$, $W_{3}=J_{-\alpha_{2}-\alpha_{3}}$ and
$W_{4}=J_{-\alpha_{1}-\alpha_{2}-\alpha_{3}}$ is the standard quadratic
$W_{4}$-algebra \cite{FL}, \cite{BG}, \cite{BBSS}.

\vspace{0.2cm}

{\bf (3b)}
{\it $A_{3}/{\cal N}_{+}^{(1,1)}\otimes U(1)\equiv V_{4}^{(1,1)}$
-algebra}: ${\cal N}_{+}^{(1,1)}=\{ E_{[\alpha ]_{1}}\}$, where $[\alpha ]_{1}$
are all positive roots, but $\alpha_{1}$,
$U(1)=\lambda_{1}\cdot H$, the constraints are
$J_{\alpha_{2}}=J_{\alpha_{3}}=1$,
$J_{\alpha_{1}+\alpha_{2}}=J_{\alpha_{2}+\alpha_{3}}=0$,
$J_{\alpha_{1}+\alpha_{2}+\alpha_{3}}=0$, $\lambda_{1}^{i}J_{i}=0$ and the
{\it g. f.} conditions are $\alpha_{2}^{i}J_{i}=\alpha_{3}^{i}J_{i}=0$,
$J_{-\alpha_{2}}=J_{-\alpha_{1}-\alpha_{2}}=0$, $J_{-\alpha_{1}}=0$. The spin 2
currents $V^{+}=J_{\alpha_{1}}$ and
$V^{-}=J_{-\alpha_{1}-\alpha_{2}-\alpha_{3}}$ are nonlocal and
$W_{3}=J_{-\alpha_{2}-\alpha_{3}}$ ($s_{3}=3$), $T=J_{-\alpha_{3}}$ ($s_{T}=2$)
are local ones. Their algebra is a nonlocal extension of the $W_{3}$-one
(\ref{18})
\begin{eqnarray}
\{ V^{+}(\sigma ),V^{-}(\sigma^{\prime})\}
&=&\partial_{\sigma^{\prime}}^{3}\delta (\sigma -\sigma^{\prime})
-T(\sigma^{\prime})\partial_{\sigma^{\prime}}\delta (\sigma -\sigma^{\prime})
-W_{3}(\sigma^{\prime})\delta (\sigma -\sigma^{\prime})
\nonumber
\\
&+&\frac{1}{3}V^{+}(\sigma )V^{-}(\sigma^{\prime})
\epsilon (\sigma -\sigma^{\prime}),
\nonumber
\end{eqnarray}
\begin{eqnarray}
\{ W_{3}(\sigma ),V^{\pm}(\sigma^{\prime})\}
&=&\mp \frac{10}{3}V^{\pm}(\sigma^{\prime})\partial_{\sigma^{\prime}}^{2}
\delta (\sigma -\sigma^{\prime})
\mp 5\partial_{\sigma^{\prime}}V^{\pm}(\sigma^{\prime})
\partial_{\sigma^{\prime}}\delta (\sigma -\sigma^{\prime})
\nonumber
\\
&\pm&\frac{1}{3}[T(\sigma^{\prime})V^{\pm}(\sigma^{\prime})
-6\partial_{\sigma^{\prime}}^{2}V^{\pm}(\sigma^{\prime})]
\delta (\sigma -\sigma^{\prime}),
\nonumber
\end{eqnarray}
\begin{eqnarray}
\{ V^{\pm}(\sigma ),V^{\pm}(\sigma^{\prime})\}
=-\frac{1}{3}V^{\pm}(\sigma )V^{\pm}(\sigma^{\prime})
\epsilon (\sigma -\sigma^{\prime}),
\label{22}
\end{eqnarray}
and the remaining $PB$ $\{ W_{3}(\sigma ),W_{3}(\sigma^{\prime})\}$ has the same
form as in (\ref{18}), but $T^{2}$, in the quadratic terms, is replaced by
$T^{2}+6V^{+}V^{-}$.

\vspace{0.2cm}

{\bf (3c)} {\it $A_{3}/{\cal N}_{+}^{(1,2)}\equiv U_{4}^{(1,2)}$
-algebra}: ${\cal N}_{+}^{(1,2)}=\{ E_{[\alpha ]_{2}}\}$, where $[\alpha ]_{2}$
are all positive roots, but $\alpha_{2}$, the constraints are
$J_{\alpha_{1}}=J_{\alpha_{3}}=1$,
$J_{-\alpha_{1}-\alpha_{2}}=J_{-\alpha_{2}-\alpha_{3}}=0$,
$J_{-\alpha_{1}-\alpha_{2}-\alpha_{3}}=0$ and the {\it g. f.} conditions are
$\alpha_{1}^{i}J_{i}=0$, $\alpha_{3}^{i}J_{i}=0$,
$J_{\alpha_{1}+\alpha_{2}}=J_{\alpha_{2}}=0$, $J_{\alpha_{2}+\alpha_{3}}=0$. The
{\it nonlocal quadratic} $U_{4}^{(1,2)}$-algebra is generated by one spin 1
current $J=\lambda_{2}^{i}J_{i}$, three local spin 2 currents
$V^{+}=J_{\alpha_{2}}$, $V^{-}=J_{-\alpha_{1}-\alpha_{2}-\alpha_{3}}$,
$T=J_{-\alpha_{1}}+J_{-\alpha_{3}}+4J^{2}$ and one nonlocal spin 2 current
$U=J_{-\alpha_{3}}-J_{-\alpha_{1}}$ \cite{GSSZ}
\begin{eqnarray}
\{ U(\sigma ),J(\sigma^{\prime})\}
=\{ V^{\pm}(\sigma ),V^{\pm}(\sigma^{\prime})\} =0,& &
\{ U(\sigma ),V^{\pm}(\sigma^{\prime})\}
=\frac{1}{2}V^{\pm}(\sigma )U(\sigma^{\prime})
\epsilon (\sigma -\sigma^{\prime}),
\nonumber
\end{eqnarray}
\begin{eqnarray}
\{ J(\sigma ),V^{\pm}(\sigma^{\prime})\}
=\mp \frac{1}{4}V^{\pm}(\sigma )\delta (\sigma -\sigma^{\prime}),& &
\{ J(\sigma ),J(\sigma^{\prime})\}
=\frac{1}{8}\partial_{\sigma}\delta (\sigma -\sigma^{\prime}),
\nonumber
\end{eqnarray}
\begin{eqnarray}
\{ V^{\pm}(\sigma ),V^{\mp}(\sigma^{\prime})\}
&=&-\frac{1}{2}\partial_{\sigma}^{3}\delta (\sigma -\sigma^{\prime})
+{\cal T}(\sigma )\partial_{\sigma}\delta (\sigma -\sigma^{\prime})
+\frac{1}{2}\partial_{\sigma}{\cal T}(\sigma )
\delta (\sigma -\sigma^{\prime})
\nonumber
\\
&-&\frac{1}{4}U(\sigma )U(\sigma^{\prime})\epsilon (\sigma -\sigma^{\prime}),
\nonumber
\end{eqnarray}
\begin{eqnarray}
\{ U(\sigma ),U(\sigma^{\prime})\}
&=&-\partial_{\sigma}^{3}\delta (\sigma -\sigma^{\prime})
+2{\cal T}(\sigma )\partial_{\sigma}
\delta (\sigma -\sigma^{\prime})
+\partial_{\sigma}{\cal T}(\sigma )\delta (\sigma -\sigma^{\prime})
\nonumber
\\
&-&[V^{+}(\sigma )V^{-}(\sigma^{\prime})+V^{+}(\sigma^{\prime})V^{-}(\sigma )]
\epsilon(\sigma -\sigma^{\prime}),
\label{23}
\end{eqnarray}
where ${\cal T}=T-4J^{2}$. As it is shown in ref \cite{GSSZ}, by choosing
another set of gauge fixing conditions, the nonlocal $U_{4}^{(1,2)}$-algebra
takes the form of the {\it rational} ($\frac{1}{U}$-terms) local algebras of
ref \cite{O'Rai2}. If one further imposes $J=\lambda_{2}^{i}J_{i}=0$ as a new
constraint, the $J$-reduced $U_{4}^{(1,2)}$-algebra
(i. e., $U_{4}^{(1,2)}/U(1)$) coincides with the nonlocal
$V_{4}^{(1,2)}$-algebra (see sec. 7 of ref \cite{GSZ1}). The main difference
with $U_{4}^{(1,2)}$ is that the spin 2 currents $V^{\pm}$ become
{\it nonlocal}, in the $V_{4}^{(1,2)}$-case.

\vspace{0.5cm}

The analysis of the above examples of $H$-reduced $SL(n,R)$-current algebras
allows to conclude that they all fit into the following {\it basic algebraic
structures}:

\vspace{0.5cm}

{\bf (A)} $W$-algebras (quadratic): {\bf (2a)}, {\bf (2e)}, {\bf (3a)} and
$W_{n}^{(2)}$ of ref \cite{Bersh}, $W_{S}^{G}$ of ref \cite{O'Rai2};

\vspace{0.1cm}

{\bf (B)} $U$-algebras (rational or nonlocal): {\bf (3c)};

\vspace{0.1cm}

{\bf (C)} $V$-albegras (nonlocal or $PF$-type): {\bf (1b)}, {\bf (2b)},
{\bf (2c)};

\vspace{0.1cm}

\noindent
and the following mixtures of {\bf (A)} and {\bf (B)} with {\bf (C)}:

\vspace{0.1cm}

{\bf (D)} $WV$-algebras (nonlocal ($PF$) quadratic): {\bf (2d)} and {\bf (3b)}
(and all $V_{n+1}^{(1,1)}$-algebras of sec. 3 of ref \cite{GSZ1});

\vspace{0.2cm}

{\bf (E)} $UV$-algebras (nonlocal ($PF$) rational):
$V_{4}^{(1,2)}=U_{4}^{(1,2)}/U(1)$.

\noindent
This observation adresses {\it the question} about the algebraic conditions that
a given set of constraints (and gauge fixing conditions) $\{ H\} \in {\cal G}$
should satisfy in order to lead to one of the above mentioned algebraic
structures ($U$, $W$, $V$, $UV$, $UW$). To answer this question, as well as
whether other families of algebras can exist, we need an {\it efficient method}
for describing all unequivalent (and irreducible) sets of first class
constraints one can impose on the currents of a given affine algebra
$\hat{\cal G}$. Given a  Lie algebra ${\cal G}_{r}$, by introducing a
grading operator\footnote{$\vec{\lambda_{n}}$ are the fundamental weights of
${\cal G}_{r}$, $\vec{\alpha}_{n}$ its simple roots, $\vec{H}$ its Cartan
subalgebra and $s_{n}$ are nonnegative integers.}
$Q_{r}^{(s)}
 = \sum_{n=1}^{r}s_{n}\frac{\vec{2\lambda_{n}}\cdot \vec{H}}{\alpha_{n}^{2}}$ we
provide it with a specific graded structure\footnote{The nonequivalent graded
structures ${\cal G}_{r}$ can have (i. e., the set of the allowed
$Q_{r}^{(s)}$), are given by the Kac theorem \cite{Kac2}; this
method was introduced in ref \cite{LS}, in the construction of the conformal
non-Abelian Toda models.}
\begin{center}
${\cal G}_{r}=\oplus_{i}{\cal G}_{\pm i}^{(s)},
\, \, \,\,\,\,
[Q_{r}^{(s)},{\cal G}_{\pm i}^{(s)}]=\pm i{\cal G}_{\pm i}^{(s)},
\, \, \,\,\,\,
[{\cal G}_{i},{\cal G}_{j}]\subset {\cal G}_{i+j}$.
\end{center}

\noindent
For each fixed $l=1,2,...,r$ (and $Q_{r}^{(s)}$), define the nilpotent
subalgebra ${\cal N}_{+}^{(l,s)}=\oplus_{i=l}{\cal G}_{i}$ and choose a generic
element $\epsilon_{+}^{(l)}\in {\cal G}_{l}$, i. e.,
$\epsilon_{+}^{(l)}
=\sum_{\alpha \in [\alpha ]_{l}}\mu_{\alpha}E_{[\alpha ]_{l}}$ where
$E_{[\alpha ]_{l}}$ are all the step operators of grade $l$ and $\mu_{\alpha}$
are arbitrary constants. Next, we consider the $\epsilon_{+}^{(l)}$-invariant
subalgebras of ${\cal G}_{0}^{(l)}=\oplus_{i=0}^{l-1}{\cal G}_{i}$ and
${\cal G}_{-}^{(l)}=\oplus_{i=l}^{r}{\cal G}_{-i}$

\begin{center}
${\cal K}_{\epsilon}^{0}(s,l)
={\rm ker ad}\epsilon_{+}^{(l)}\cap {\cal G}_{0}^{(l)}
=\{ g_{0}^{0}\in {\cal G}_{0}^{(l)}:[\epsilon_{+}^{(l)},g_{0}^{0}]=0\}$, 
\end{center}

\begin{center}
${\cal K}_{\epsilon}^{-}(s,l)
={\rm ker ad}\epsilon_{+}^{(l)}\cap {\cal G}_{-}^{(l)}
=\{ g^{-}\in {\cal G}_{-}^{(l)}:[\epsilon_{+}^{(l)},g^{-}]=0\}$.
\end{center}

Finally, we define the ``constraint'' subalgebra as
${\cal H}_{c}^{(l,s)}(\epsilon_{+}^{(l)})
={\cal N}_{+}^{(l)}(\epsilon )\oplus {\cal H}_{0}^{(l)}$, where
${\cal H}_{0}^{(l)}\subset {\cal K}_{\epsilon}^{0}(s,l)$ denotes those
subalgebras of ${\cal K}_{\epsilon}^{0}$, which elements (i. e., the currents
belonging to ${\cal H}_{0}^{(l)}$) are constrained to zero;
${\cal N}_{+}^{(l)}(\epsilon )$ caries the information about the constraints we
are imposing on the currents from ${\cal N}_{+}^{(l)}$, namely all the elements
of the subalgebra ${\cal N}_{+}^{(l+1)}\subset {\cal N}_{+}^{(l)}$ are zero and
the elements of ${\cal G}_{l}$, which are constrained to be constants
$\mu_{\alpha}\neq 0$, are collected in
$\epsilon_{+}^{(l)}=\sum_{\alpha}\mu_{\alpha}E_{\alpha}$ (all the remaining
${\cal G}_{l}$ elements are zero). In this language, {\it the problem of the
classification} of the allowed set of constraints
${\cal H}_{c}^{(l,s)}(\epsilon )$ reads as follows: for each fixed $Q_{r}^{(s)}$
and fixed grade $l$ (say $l=1$), to make a list of all the nonequivalent choises
of the $\epsilon_{+}^{(l)\prime}$s. One can further organizes the different sets
of $\epsilon_{+}^{(l)\prime}$s ($l$ and $Q_{r}^{(s)}$ fixed) in {\it families}
(${\cal K}_{\epsilon}^{0}(s,l)$, ${\cal K}_{\epsilon}^{-}(s,l)$), according to
their {\it invariant subalgebras}. For example, the family
($\lambda_{i}\cdot H$, ${\rm \O}$) is characterized by the conditions: (a) 
$\mu_{\alpha}=0$, for the $\vec{\alpha^{i}}\cdot \vec{\lambda_{i}}\neq 0$
($\alpha \in [\alpha ]_{l}$), and (b) $i=1$ or $i=r$ (for the $l=1$ case), in
order to have ${\cal K}_{\epsilon}^{-}(s,1)={\rm \O}$. We call {\it equivalent}
the sets of constraints (and gauge fixing conditions) which can be obtained from
each other by certain discrete transformations from the Weyl group of
${\cal G}_{r}$. As it is shown in sec. 8 of ref \cite{GSZ1} (for the grade
$l=1$), they give rise to the same ${\cal H}_{c}^{(l,s)}(\epsilon )$-reduced
${\cal G}_{r}$-algebra. Therefore, it is sufficient to consider only one
representative of such ``Weyl families'' of constraints. The problem of the
{\it irreducibility} is more delicate. Depending on our choice of
$\epsilon_{+}^{(l)\prime}$s, it might happens that the
${\cal G}_{r}/{\cal H}_{c}^{(l,s)}(\epsilon )$-algebra splits into two
(or more) mutualy commuting algebras \cite{GSSZ}. This is the case when one
takes, for example $\mu_{\alpha}=0$, for all $E_{\alpha}$ that contains the
simple root $\alpha_{i}$ (i. e., $E_{\alpha_{i}}$,
$E_{\alpha_{i}+\alpha_{i+1}}$, $E_{\alpha_{i-1}+\alpha_{i}}$,
$E_{\alpha_{i-1}+\alpha_{i}+\alpha_{i+1}}$ etc).

The organization of the constraints in the families
(${\cal K}_{\epsilon}^{0}(s,l)$, ${\cal K}_{\epsilon}^{-}(s,l)$) simplifies the
derivation of the
${\cal G}_{r}^{(l,s)}(\epsilon ,{\cal H}_{0}^{(l)})
={\cal G}_{r}/{\cal H}_{c}^{(l,s)}(\epsilon )$-algebras (i. e., the
calculation of the corresponding Dirac brackets). Depending on the algebraic
data $\{ {\cal G}_{r},Q^{(s)},l,\epsilon_{+}^{(l)}\}$, which defines
${\cal H}_{c}^{(l,s)}(\epsilon )$, one can {\it classify} all the
${\cal G}_{r}^{(l,s)}(\epsilon ,{\cal H}_{0}^{(l)})$-algebras in the following
$\{ {\cal H}_{0}^{(l)},{\cal K}_{\epsilon}^{-}(s,l)\}$-families of algebras:

THEOREM. Given ${\cal G}_{r}$ and the graded structure
($Q_{r}^{(s)}$, $l$, $\epsilon_{+}^{(l)}$), which define the constraints
subalgebra
${\cal H}_{c}^{(l,s)}(\epsilon )\subset {\cal G}_{r}$. Each
${\cal H}_{c}^{(l,s)}(\epsilon )$-reduced ${\cal G}_{r}$-current algebra
${\cal G}_{r}^{(l,s)}(\epsilon ,{\cal H}_{0}^{(l)})$ belongs to one of the
following five types of extended Virasoro algebras:

(1) $W$-algebras, when ${\cal H}_{0}^{(l)}={\rm \O}$ (or
${\cal H}_{0}^{(l)}\neq {\rm \O}$ but
$[{\cal H}_{0}^{(l)},{\cal G}_{0}^{\pm}]=0$) and
${\cal K}_{\epsilon}^{-}={\rm \O}$ (${\cal G}^{\pm}_{0}$ are the $\pm$ step
operators of grade 0);

(2) $U$-algebras, when ${\cal H}_{0}^{(l)}={\rm \O}$ (or
${\cal H}_{0}^{(l)}\neq {\rm \O}$ but
$[{\cal H}_{0}^{(l)},{\cal G}_{0}^{\pm}]=0$) and
${\cal K}_{\epsilon}^{-}\neq {\rm \O}$; dim ${\cal K}_{\epsilon}^{-}$ is the
number of the nonlocal currents
(or of the ``rational currents'' of ref \cite{O'Rai2});

(3) $V$-algebras, when ${\cal H}_{0}^{(l)}\neq {\rm \O}$ and
${\cal H}_{0}^{(l)}=U(1)^{r}$ or
$U(1)^{r-1}=\{ \oplus_{i=2}^{r}\lambda_{i}\cdot H\}$ or
$\{ \oplus_{i=1}^{r-1}\lambda_{i}\cdot H\}$,
${\cal K}_{\epsilon}^{-}={\rm \O}$; the case $\epsilon_{+}^{(l)}=0$,
${\cal H}_{0}^{(l)}=U(1)^{r}$, which also lead to $V$-algebras, has to be
{treated separately (see ref \cite{GSSZ});

(4) $VW$-algebras, when
${\cal H}_{0}^{(l)}\neq \{ {\rm \O},\, \, U(1)^{r},\, \, U(1)^{r-1}\}$
and ${\cal K}_{\epsilon}^{-}={\rm \O}$; the case
$[{\cal H}_{0}^{(l)},{\cal G}_{0}^{\pm}]=0$ (${\cal G}_{0}^{\pm}$ are the $\pm$
step operators of grade zero) has to be excluded, since it gives rise to
$W$-algebras;

(5) $VU$-algebras, when
${\cal H}_{0}^{(l)}\neq \{ {\rm \O},\, \, U(1)^{r},\, \,
\oplus_{i=2}^{r}\lambda_{i}\cdot H ,\, \,  
\oplus_{i=1}^{r-1}\lambda_{i}\cdot H\}$,
and ${\cal K}_{\epsilon}^{-}\neq {\rm \O}$; in this case, again
$[{\cal H}_{0}^{l},{\cal G}_{0}^{\pm}]\neq 0$.

The algebraic conditions that separate $W$- from the $U$-algebras are given in
ref \cite{O'Rai2}. The equivalence of the {\it rational} $U$-algebras, to
certain {\it nonlocal} algebras, and the explicit form of the gauge
transformations, from the ``rational'' gauge fixing conditions to ``nonlocal''
gauge fixing conditions, is demonstrated in ref \cite{GSSZ}. The proof of this
theorem, for the generic $Q_{r}^{(s)}$ grade one ($l=1$) case \cite{GSSZ}, is
based on the analysis of the properties of the inverse matrix $\Delta_{ij}^{-1}$
of the constraints and the gauge fixing conditions. The {\it origin of the
nonlocal terms} in the $V$- , $VW$- and $VU$-algebras, are the
${\cal H}_{0}^{(l)}$ constraints and their gauge fixing$^{\prime}$s. Their
$PB^{\prime}$s are always in the form
$\{ J_{i}(\sigma ),J_{j}(\sigma^{\prime})\}
=k\delta_{ij}\partial_{\sigma}\delta (\sigma -\sigma^{\prime})$ or
$\{ J_{-\alpha_{i}}(\sigma ),J_{\alpha_{i}}(\sigma^{\prime})\}
=k\partial_{\sigma}\delta (\sigma -\sigma^{\prime})$. Their contributions to
$\Delta_{ij}^{-1}$ are the nonlocal $\epsilon (\sigma -\sigma^{\prime})$-terms.

The explicit form of each ${\cal G}_{r}^{(l,s)}(\epsilon ,{\cal H}_{0}^{(l)})$-
algebra (from a given class $U$, $V$, $VW$ etc) indeed depends on the algebra
${\cal G}_{r}$ and on the choice of $\epsilon_{+}^{(l)}$ and
${\cal H}_{0}^{(l)}$ as one can see from our Examples 1, 2 and 3.
The full algebraic structure (all explicit $PB^{\prime}$s) is known in the case
of the $W_{n}$-algebras \cite{FL} and of the simplest $A_{r}$-family of
$VW$-algebras ($V_{r+1}^{(1,1)}$-algebras \cite{GSZ1}) defined by
$Q=\sum_{i=l}^{r}\lambda_{i}\cdot H$, $l=1$,
$\epsilon_{+}^{(1)}=\sum_{i=2}^{r}E_{\alpha_{i}}$,
${\cal H}_{0}^{(1)}=\{ \lambda_{1}\cdot H\}$. Various examples of the $U$- $V$-
and $VU$-algebras ($V_{(n,m)}$) have been constructed by Bilal \cite{Bil2}, by
calculating the second Gelfand-Dikii brackets, associated with certain matrix
differential operators.

\section{Quantum $V$-algebras}

The classification of all the classical ($PB^{\prime}$s) extensions of the
Virasoro algebra is an important step forward the classification of the
universality classes in two dimensions. The complete solution of this problem
requires, however, the knowledge of the exact critical exponents, i. e., we need
to know the {\it h. w.} representation of the corresponding {\it quantum} $W$-,
$U$-, $V$- and $VW$-, $VU$-algebras. The quantization of the classical
$W$-algebras is a rather well understood problem. It consists in replacing the
currents functions $T$, $W_{n}$ by currents operators $\hat{T}$, $\hat{W_{n}}$,  acting
on some Hilbert space, and their $PB^{\prime}$s $\{ a,b\}$ by the commutators
$-\frac{\imath}{\hbar}[a,b]$. The only changes that occur in this procedure are
the new ({\it quantum corrections}) coefficients in front of the central term
$\delta^{(s)}(\sigma )$ and those of the quadratic terms. Another option is to
start with the quantum current algebra ${\cal G}_{r}$ and to implement the
operators constraints ${\cal H}_{c}^{(l,s)}(\epsilon )$ on it, following the
methods of the {\it quantum Hamiltonian reduction} \cite{BO}. The advantage of
this method is that it provides a simple way of deriving the $W$-algebra
{\it h. w.} representations from the {\it h. w.} representations of the
${\cal G}_{r}$-current algebra. The specific nonlocal terms
$V^{+}(\sigma )V^{-}(\sigma^{\prime})\epsilon (\sigma -\sigma^{\prime})$ that
appears in the $V$- (and $VW$-, $VU$-) algebras, as well as the nonlocal nature
of the part of the currents ($V_{i}^{\pm}$), are the main obstacle to the
construction of the corresponding {\it quantum} $V$-algebras. It turns out
\cite{GSZ1}, \cite{GSZ2}
that their quantization require deep changes in the classical algebraic
structure (\ref{13}), (\ref{20}), (\ref{21}), namely: (a) renormalization
of the bare spins of the nonlocal currents (say for $V_{n+1}^{(1,1)}$,
$s_{cl}^{\pm}=\frac{n+1}{2}$ goes to
$s_{q}^{\pm}=\frac{n+1}{2}\left( 1-\frac{1}{2k+n+1}\right)$); (b) the
quantum counterpart of the $PB^{\prime}$s of the $V^{\pm \prime}$s charges
appears to be specific $PF$-type commutators, similar to eqn. (\ref{10});

\noindent
(c) breaking of the global $U(1)$ symmetry, to some discrete group
$Z_{2k+n}$.

The fact that all the complications in the quantization of the classical
$V$- and $VW$-algebras are coming from the ${\cal H}_{0}^{(l)}$-constraints
suggests the following {\it strategy}: relax the
${\cal H}_{0}^{(l)}$-constraints (i. e., leave the currents
$\lambda_{a}^{i}J_{i}\in {\cal H}_{0}^{(l)}$ unconstrained) and consider the
corresponding local ``{\it intermidiate}'' $W$-algebra, generated by the
$V$- (or $VW$-) algebra currents (which are all local now) and the additional
spin one ${\cal H}_{0}^{(l)}$-currents. Since all the currents are local, the
quantization of this algebra is similar to the one of the $W_{n+1}$- or
$W_{n+1}^{(l)}$-algebras \cite{FL}, \cite{BO}. The problem we address here is
the following: {\it Given the quantum $W$-algebra and its {\it h. w.}
representations, to derive the quantum $V=W/{\cal H}_{0}^{(l)}$-algebra and its {\it h. w.}
representations by implementing the (operator) constraint
${\cal H}_{0}^{(l)}\approx 0$}. The method we are going to use is an appropriate
generalization of the derivation of the $Z_{N}$ parafermionic algebra \cite{ZF1}
from the affine $SU(2)$-one (or $SL(2,R)$, for the noncompact $PF^{\prime}$s),
by imposing the constraint $J_{3}(z)\approx 0$.

{\bf Example 3.1}. {\bf Quantization of the $PF$-algebra}. Following the
arguments of ref \cite{ZF1}, we define the quantum (compact) $V_{2}$-algebras as
$V_{2}=\{ SU(2)_{k},J_{3}(z)=0\}$. Therefore, the $V_{2}$-generators
$\psi^{\pm}$ have to represent the
$J_{3}=\sqrt{\frac{k}{2}}\partial \phi$ -independent part of the
$\hat{SU}(2)_{k}$-ones, i. e.,
\begin{eqnarray}
J^{\pm}=\psi^{\pm}\exp (\mp \alpha \phi ),
\ \
T=T_{V}+\frac{1}{2}(\partial \phi )^{2},
\ \
J_{3}(z_{1})\psi^{\pm}(z_{2})=O(z_{12}),
\nonumber
\end{eqnarray}
\begin{eqnarray}
\phi (z_{1})\phi (z_{2})=-\ln (z_{12})+O(z_{12}).
\label{24}
\end{eqnarray}
Taking into account the $SU(2)$ $OPE^{\prime}$s
\begin{eqnarray}
J_{3}(z_{1})J^{\pm}(z_{2})=\pm \frac{\imath}{z_{12}}J^{\pm}(z_{2})+O(z_{12})
\label{25}
\end{eqnarray}
and eqn. (\ref{24}), we find $\alpha =\imath \sqrt{\frac{2}{k}}$ and, as a
consequence, the spins of $\psi^{\pm}$ are $s^{\pm}=1-\frac{1}{k}$ (we have used
that $s_{J^{\pm}}=1$). Finally, eqns. (\ref{24}) and (\ref{25}) lead to the
following $V_{2}$-algebra $OPE^{\prime}$s
\begin{eqnarray}
\psi^{\pm}(z_{1})\psi^{\pm}(z_{2})&=&z_{12}^{-\frac{2}{k}}\psi^{\pm}_{(2)}(z_{2})
+O(z_{12}),
\nonumber
\\
\psi^{+}(z_{1})\psi^{-}(z_{2})
&=&z_{12}^{\frac{2}{k}}\left( \frac{k}{z_{12}^{2}}+(k+2)T_{V}+O(z_{12})\right) ,
\label{26}
\end{eqnarray}
which are nothing, but the $PF$-algebra $OPE^{\prime}$s (\ref{9}), with $k=N$
and $\psi_{1}^{\pm}=\frac{1}{\sqrt{k}}\psi^{\pm}$. Although the $V_{2}$-algebra
(\ref{26}) is, by construction, the quantum version of the classical
$PB^{\prime}$s $PF$-algebra (\ref{17}), the discrepancy between the spins
$s^{\pm}=1-\frac{1}{k}$ and $s_{V^{\pm}}=1$ requires a more precise definition
of the relation of algebras (\ref{26}) and (\ref{17}). The exact {\it statement}
is as follows: let $V^{\pm}=\frac{1}{k}\psi^{\pm}$ and the $V^{\pm}$
$PB^{\prime}$s are defined as a certain limit of the $OPE^{\prime}$s (\ref{26}):
\begin{eqnarray}
\{ V^{a}(z_{1}),V^{b}(z_{2})\} =\lim_{k\rightarrow \infty}\frac{k}{2\pi \imath}
[V^{a}(z_{1})V^{b}(z_{2})-V^{a}(z_{2})V^{b}(z_{1})]
\label{27}
\end{eqnarray}
($a,b=\pm$). Then, the $k\rightarrow \infty$ limit of the $OPE^{\prime}$s
(\ref{26}) reproduces the $PB^{\prime}$s (\ref{17}). The proof is
straightforward. Applying twice the $OPE^{\prime}$s (\ref{26}), we find
\begin{eqnarray}
z_{12}^{\frac{2}{k}}\{ V^{\pm}(z_{1})V^{\pm}(z_{2})-V^{\pm}(z_{2})V^{\pm}(z_{1})
e^{-\frac{2\pi \imath}{k}\epsilon}(z_{12})\}
&=&\frac{1}{k^{2}}O(z_{12}),
\nonumber
\end{eqnarray}
\begin{eqnarray}
z_{12}^{-\frac{2}{k}}\{ V^{-}(z_{1})V^{+}(z_{2})&-&V^{+}(z_{2})V^{-}(z_{1}) 
e^{\frac{2\pi \imath}{k}\epsilon}(z_{12})\} =
\frac{k+2}{k^{2}}O(z_{12})
\nonumber
\\
&+&\frac{1}{k}\left( \frac{1}{z_{12}^{2}+\imath 0}
-\frac{1}{z_{21}^{2}+\imath 0}\right) ,
\label{28}
\end{eqnarray}
where the identity
$\imath \pi \epsilon (z_{12})\equiv \ln \frac{z_{12}+\imath 0}{z_{21}+\imath 0}$
has been used. The $k\rightarrow \infty$ limit of eqns. (\ref{28}) reproduces
exactly the classical $PF$-algebra\footnote{The noncompact case
$SL(2,R)/U(1)$ corresponds to the change
$\phi \rightarrow \imath \phi$, which turns out to be equivalent to the
$k\rightarrow -k$ one, in the $OPE^{\prime}$s, spins etc.} (\ref{17}). The
conclusion is that the nonlocal $PB^{\prime}$s-algebra (\ref{17}) is a
semiclassical limit ($k\rightarrow \infty$) of the $PF^{\prime}$s
$OPE^{\prime}$s (\ref{26}). As we have seen, the quantization requires
renormalization of the spins $s_{q}=s_{cl}-\frac{1}{k}$ of the nonlocal currents
$V^{\pm}$. Therefore, the $PB^{\prime}$s (\ref{17}) have to be replaced by the
$PF$-commutators (\ref{10}) and for $k$-positive integers, the classical global
$Z_{2}\otimes U(1)$-symmetry is broken to $Z_{2}\times Z_{k}$, in the quantum
theory.

The structure of the classical $V_{3}^{(1,1)}$ $PB^{\prime}$s algebra (\ref{13})
is quite similar to the $PF$-one (\ref{17}). An importante difference is that in
its derivation from the classical $SL(3,R)$ (see our example 2b), one
has to impose, together with the ${\cal H}_{0}^{1}$-type ($PF$) constraint
$\vec{\lambda_{1}}\cdot \vec{J}=0$, two more constraints, on the nilpotent
subalgebra ${\cal N}_{+}^{(1)}$: $J_{\alpha_{2}}=1$ and
$J_{\alpha_{1}+\alpha_{2}}=0$. In order to demonstrate how this type of (purely
$W$-) constraints are treated, in the frameworks of the quantum Hamiltonian
reduction, we consider the simplest example of such reduction: the
${\cal N}_{+}$-reduced $SL(2,R)$ ($J_{\alpha}=1$) which gives rise
to the Virasoro algebra (example 1a).

\vspace{0.2cm}

{\bf Example 3.2}. {\bf Virasoro algebra {\it h. w.} representation from the
$SL(2,R)$-ones} \cite{BO}. The implementation of the constraint
$J_{\alpha}=1$ as an operator identity on the $SL(2,R)_{k}$-space of states
${\cal H}_{A_{1}}^{(k)}$ requires an introduction of a pair of fermionic ghosts
($b(z)$, $c(z)$) of spins (0,1) and of the larger space of states
${\cal H}_{A_{1}}^{(k)}\otimes {\cal H}_{b,c}$. The reduced representation space
of the constrained system $\{ A_{1}/{\cal N}_{+}\}$ can be defined by
means of the $BRST$ operator

\begin{center}
$Q_{BRST}=\oint [J_{\alpha}(z)-1]c(z)dz,$\, \,
$Q_{BRST}^{2}=0$,
\end{center}

\noindent
as $Q_{BRST}$-invariant
states-$|\psi >\in {\cal H}_{A_{1}}^{(k)}\otimes {\cal H}_{b,c}$
($Q_{BRST}|\psi >=0$), which are not $Q_{BRST}$-exact, i. e.,
$|\psi >\neq Q_{BRST}|\ast >$. The statement is that this $BRST$-cohomology
$H_{Q_{BRST}}({\cal H}_{A_{1}}^{(k)}\otimes {\cal H}_{b,c})
={\rm ker}Q/{\rm Im}Q$ is isomorphic to the irreducible representation space
${\cal H}_{Vir}^{(k)}(\equiv {\cal H}_{A_{1}}/{\cal N}_{+})$ of the
Virasoro algebra \cite{BO}. To make the constraints condition $J_{\alpha}=1$
consistent with the conformal invariance, we have to improve the
$SL(2,R)$-Sugawara stress-tensor, in such a way that
$s_{imp}(J_{\alpha})=0$
\begin{eqnarray}
T_{impr}=\frac{1}{k+2}:J^{a}(z)J^{a}(z):+\partial J_{3}.
\nonumber
\end{eqnarray}
Therefore, the new central charge is $c_{impr}=\frac{3k}{k+2}-6k$. Taking into
account the contribution $c_{gh}=-2$, of the ghost stress-tensor
$T_{bc}=(\partial b)c$, we find that the total central charge is
$c_{tot}=13-6\left( \frac{1}{k+2}+k+2\right)$. Since the dimension of the
$\hat{SL}(2,R)_{k}$ representation of weight $\vec{\Lambda}$ is
\begin{eqnarray}
\Delta_{\Lambda}
=\frac{1}{2(k+2)}\vec{\Lambda}\cdot (\vec{\Lambda}+2\vec{\alpha})
\nonumber
\end{eqnarray}
the improved dimensions are found to be
\begin{eqnarray}
\Delta_{\Lambda}^{impr}
=\frac{1}{2(k+2)}\vec{\Lambda}\cdot (\vec{\Lambda}+2\vec{\alpha})
-\vec{\alpha}\cdot \vec{\Lambda}.
\label{29}
\end{eqnarray}
An important observation of ref \cite{BO} is that the {\it h. w.} states of
the reduced space ${\cal H}_{Vir}^{(k)}$ are of levels $k+2=\frac{m}{m+1}$,
where $m=3,4,...$, and weights
\begin{eqnarray}
\vec{\Lambda}=[(1-p)(k+2)-(1-q)]\vec{\alpha},
\nonumber
\end{eqnarray}
with $1\leq p\leq m-1$, $1\leq q\leq p$. Therefore, $c_{tot}=1-\frac{6}{m(m+1)}$
and $\Delta_{\Lambda}^{impr}=\Delta_{p,q}$, i. e., the ${\cal H}_{Vir}^{(k)}
=H_{Q_{BRST}}({\cal H}_{A_{1}}^{(k)}\otimes {\cal H}_{b,c})$ (for the above
values of the levels, and the $A_{1}$-weight $\vec{\Lambda}$) coincides with the
space of the {\it h. w.} unitary representations (\ref{3}) of the Virasoro
algebra.

\vspace{0.2cm}

{\bf Example 3.3}. {\bf Quantum $V_{3}^{(1,1)}$-algebra}. As it was pointed out in
ref \cite{GSZ1}, \cite{GSZ2}, the intermidiate $W_{3}^{(1,1)}$-algebra is Weyl
equivalent ($w_{\alpha_{1}}$) to the Bershadsky-Polyakov algebra $W_{3}^{(2)}$. The
improved stress-tensor is given by
\begin{eqnarray}
T^{impr}=\frac{1}{k+3}:J^{q}(z)J^{q}(z):
-(\lambda_{2}-\frac{1}{2}\lambda_{1})^{i}\partial J_{i}
\label{30}
\end{eqnarray}
and we have to introduce the following two pair of ghosts: ($b,c$) and
($\phi ,\phi^{+}$), of spins (0,1) and ($\frac{1}{2},\frac{1}{2}$).
Constructions, similar to the ones in { Example 3.2}, allow to derive the
$W_{3}^{(1,1)}$ central charge

\begin{center}
$c_{W_{3}^{(1,1)}}=\frac{8k}{k+3}-6k-1$
\end{center}

\noindent
and the dimensions $\Delta_{\vec{r},\vec{s}}$ and $U(1)$ charges $q_{r,s}$ of
its {\it h. w.} representations ($NS$-sector) \cite{Bersh} read as
\begin{eqnarray}
\Delta_{\vec{r},\vec{s}}^{W}
&=&\frac{1}{2(k+3)}\vec{\Lambda}_{r,s}\cdot (\vec{\Lambda}_{r,s}+2\vec{\beta})
-\vec{\beta}\cdot \vec{\Lambda}_{r,s},
\nonumber
\\
q_{\vec{r},\vec{s}}
&=&\frac{1}{3}\left[ 2\frac{p}{q}(r_{1}-r_{2})-(s_{1}-s_{2})\right] ,
\label{31}
\end{eqnarray}
where $\beta =\lambda_{2}-\frac{1}{2}\lambda_{1}$, with $\vec{\Lambda}_{r,s}$
representing the weights of the following specific level $k+3=2\frac{p}{q}$
representations of $SL(3,R)_{k}$
\begin{eqnarray}
\vec{\Lambda}_{r,s}=\sum_{i=1}^{2}\vec{\lambda}_{i}[(1-r_{i})(k+3)-(1-s_{i})],
\nonumber
\end{eqnarray}
where $1\leq s_{i}\leq 2p-1$, $1\leq r_{i}\leq q$. The quantum
$W_{3}^{(1,1)}$-algebra is generated by one spin $s=1$ $J(z)$, two
spin $s=\frac{3}{2}$ $G^{\pm}(z)$ and one spin $s=2$ $T(z)$ currents, with
$OPE^{\prime}$s \cite{Bersh}
\begin{eqnarray}
J(z_{1})J(z_{2})=\frac{2k+3}{3z_{12}^{2}}+O(z_{12}),& &
J(z_{1})G^{\pm}(z_{2})=\pm \frac{1}{z_{12}}G^{\pm}(z_{2})+O(z_{12}),
\nonumber
\end{eqnarray}
\begin{eqnarray}
G^{+}(z_{1})G^{-}(z_{2}) &=&\frac{(k+1)(2k+3)}{z_{12}^{3}}
+3\frac{k+1}{z_{12}^{2}}J(z_{2})
\nonumber
\\
&+&\frac{1}{z_{12}}\left[ 3J^{2}(z_{2})-(k+3)T(z_{2})
+3\frac{k+1}{2}\partial J(z_{2})\right] +O(z_{12}).
\nonumber
\end{eqnarray}
\begin{eqnarray}
G^{\pm}(z_{1})G^{\pm}(z_{2})=O(z_{12}),
\label{32}
\end{eqnarray}
According to the definition of the $V_{3}^{(1,1)}$-algebra
$V_{3}^{(1,1)}=\{ W_{3}^{(1,1)};J=0\}$, its generators $V^{\pm}(z)$ and $T_{V}$
have to commute with $J(z)$, i. e.,
\begin{eqnarray}
J(z_{1})V^{\pm}(z_{2})=J(z_{1})T_{V}(z_{2})=O(z_{12}).
\label{33}
\end{eqnarray}
Therefore, $V^{\pm}$, $T_{V}$ are related to the
$J=\sqrt{\frac{2k+3}{3}}\partial \phi$-independent parts of the
$W_{3}^{(1,1)}$-currents
\begin{eqnarray}
G^{\pm}=V^{\pm}\exp (\pm a\phi ),\, \, 
T_{W}=T_{V}+\frac{1}{2}(\partial \phi )^{2},\, \,
\phi (z_{1})\phi (z_{2})=\ln (z_{12})+O(z_{12}).
\label{34}
\end{eqnarray}
As a consequence of eqns. (\ref{33}) and (\ref{34}), we get
$a=\sqrt{\frac{3}{2k+3}}$, and for the spins of the quantum currents
$V^{\pm}\left( s_{cl}=\frac{3}{2}\right)$, we obtain
$s_{q}^{\pm}=\frac{3}{2}\left( 1-\frac{1}{2k+3}\right)$. The
$W_{3}^{(1,1)}-OPE^{\prime}$s (\ref{32}), and eqn. (\ref{34}), lead to the
following $OPE^{\prime}$s for $V^{\pm}$ and $T_{V}$ ($k\neq -3,-\frac{3}{2},-1$)
\begin{eqnarray}
&\frac{}{}&V^{+}(z_{1})V^{-}(z_{2})
=z_{12}^{\frac{3}{2k+3}}\left[ \frac{(2k+3)(k+1)}{z_{12}^{3}}
-\frac{k+3}{z_{12}}T_{V}(z_{2})+O(z_{12})\right] ,
\nonumber
\\
&\frac{}{}&T_{V}(z_{1})V^{\pm}(z_{2})=\frac{s^{\pm}_{q}}{z_{12}^{2}}V^{\pm}(z_{2})
+\frac{1}{z_{12}}\partial V^{\pm}(z_{2})+O(z_{12}),
\nonumber
\\
&\frac{}{}&V^{\pm}(z_{1})V^{\pm}(z_{2})=z_{12}^{-\frac{3}{2k+3}}V^{\pm}_{(2)}(z_{2})
+O(z_{12}),
\label{35}
\end{eqnarray}
which define the {\it quantum} $V_{3}^{(1,1)}$-algebra. The $T_{V}(1)T_{V}(2)$
$OPE$ has the standard form (\ref{4}), of the Virasoro $OPE^{\prime}$s, with
central charge $c_{V}=-6\frac{(k+1)^{2}}{k+3}$. The $V_{3}^{(1,1)}$-algebra
(\ref{35}) has a structure similar to the $PF$-one (\ref{9}), and for $L=2k+3$
positive integers ($L>3$), the $OPE^{\prime}$s (\ref{35}) involves more currents
$V^{\pm}_{l}$ ($l=1,2,...,L-1$) of spins $s^{\pm}_{l}=\frac{3l}{2L}(L-l)$.
Introducing the (Laurent) mode expansion for the
currents\footnote{$\phi_{s}^{\eta}(0)$ denotes certain Ramond
($\eta =\frac{1}{2}$, $s$-odd) and Neveu-Schwartz ($\eta =0$, $s$-even) fields,
where $s=1,2,...,L-1$.}$V^{\pm}$
\begin{eqnarray}
V^{\pm}(z)\phi_{s}^{\eta}(0)
=\sum_{m=-\infty}^{\infty}z^{\pm \frac{3s}{2L}+m-1\mp \eta}
V^{\pm}_{-m\pm \eta -\frac{1}{2}+3\frac{1\mp s}{2L}}\phi_{s}^{\eta}(0),
\nonumber
\end{eqnarray}
we derive, from (\ref{35}), the following ``commutation relations'' (of
$PF$-type) for the $V_{3}^{(1,1)}(L)$-algebra ($|L|>3$)
\begin{eqnarray}
 \sum_{p=0}^{\infty}C^{p}_{(-\frac{3}{L})}
\left( V^{+}_{-3\frac{s+1}{2L}+m-p-\eta +\frac{1}{2}}
V^{-}_{3\frac{s+1}{2L}+n+p+\eta -\frac{1}{2}}
+V^{-}_{-3\frac{1-s}{2L}+n-p+\eta -\frac{1}{2}}
V^{+}_{3\frac{1-s}{2L}+m+p-\eta +\frac{1}{2}}\right)
\nonumber
\er
\br
 =\frac{1}{2}(L+3)\left[ -L_{m+n}+\frac{(L-1)L}{2(L+3)}
\left( \frac{3s}{2L}+n+\eta \right)
\left( \frac{3s}{2L}+n+\eta -1\right) \delta_{m+n,0}\right] 
\label{36}
\end{eqnarray}
where $C_{(M)}^{p}=\frac{\Gamma (p-M)}{p!\Gamma (-M)}$, with
$m,n=0,\pm 1,\pm 2,...$, and
\begin{eqnarray}
\sum_{p=0}^{\infty}C^{p}_{(\frac{3}{L})}
\left( V^{\pm}_{3\frac{3\mp s}{2L}-p+m+\eta -\frac{1}{2}}
V^{\pm}_{3\frac{1\mp s}{2L}+p+n+\eta -\frac{1}{2}}
-V^{\pm}_{3\frac{3\mp s}{2L}-p+n+\eta -\frac{1}{2}}
V^{\pm}_{3\frac{1\mp s}{2L}+p+m+\eta -\frac{1}{2}}\right )=0.
\label{37}
\end{eqnarray}
In the particular cases, when $L=2,3$, the $OPE^{\prime}$s $V^{\pm}V^{\pm}$ have
also a pole, which makes eqn. (\ref{37}) nonvalid. The simplest example of such
$V_{3}^{(1,1)}$-algebra, for $L=2$, is spanned by $V^{\pm}$ of
$s^{\pm}=\frac{3}{4}$ and $T_{V}$, only. The relations (\ref{37}) are now
replaced by
\begin{eqnarray}
\sum_{p=0}^{\infty}C^{p}_{\left( \frac{1}{2}\right)}
(V^{-}_{-p+m+\eta -\frac{3}{4}}V^{-}_{p+n+\eta -\frac{5}{4}}
+V^{-}_{-p+n+\eta -\frac{3}{4}}V^{-}_{p+m+\eta-\frac{5}{4}})
=\delta_{m+n+2\eta ,0},
\nonumber
\end{eqnarray}
and by the similar one, for $V^{+}V^{+}$. As in the $PF$-case, one can easily
verify that certain limits of the $OPE^{\prime}$s (\ref{35}) reproduces the
classical $PB^{\prime}$s $V_{3}^{(1,1)}$-algebra (\ref{13}).

The relations (\ref{34}), between $W_{3}^{(1,1)}$ and $V_{3}^{(1,1)}$ currents,
lead to the following form for the $W_{3}^{(1,1)}$-vertex operators
$\phi^{W}_{(r_{i},s_{i})}(z)$ in terms of the $V_{3}^{(1,1)}$-ones
$\phi^{V}_{(r_{i},s_{i})}(z)$ and the free field $\phi$
\begin{eqnarray}
\phi^{W}_{(r_{i},s_{i})}=\phi^{V}_{(r_{i},s_{i})}
\exp \left[ q_{(r_{i},s_{i})}\sqrt{\frac{3}{L}}\phi \right] .
\label{38}
\end{eqnarray}
The construction (\ref{38}) is a consequence of eqns. (\ref{34}), of the
following $OPE^{\prime}$s
\begin{eqnarray}
T^{W}(z_{1})\phi^{W}(z_{1})
&=&\frac{\Delta^{W}_{r,s}}{z_{12}^{2}}\phi^{W}_{(r,s)}(z_{2})
+\frac{1}{z_{12}}\partial \phi^{W}_{(r,s)}(z_{2})+O(z_{12}),
\nonumber
\\
J(z_{1})\phi^{W}(z_{2})
&=&\frac{q_{r,s}}{z_{12}}\phi^{W}_{(r,s)}(z_{2})+O(z_{12}),
\nonumber
\end{eqnarray}
and of the fact that $\phi^{V}_{(r,s)}$ are $J$-neutral, i. e.,
$J(z_{1})\phi^{V}_{(r,s)}(z_{2})=O(z_{12})$. Finally, we realize that the
dimensions $\Delta^{V}_{(r,s)}$ of the $V_{3}^{(1,1)}$ primary fields
$\phi^{V}_{(r,s)}$ are related to the $\phi^{W}_{(r,s)}$ dimensions and
charges, given by eqns. (\ref{31}), as follows
\begin{eqnarray}
\Delta^{V}_{(r,s)}=\Delta^{W}_{(r,s)}-\frac{3}{2L}q^{2}_{(r,s)}.
\label{39}
\end{eqnarray}
Taking into account the explicit values of $\Delta^{W}_{(r,s)}$ and $q_{(r,s)}$
(\ref{31}), for the class of ``completely degenerate'' {\it h. w.}
representations of $W_{3}^{(1,1)}$ ($L+3=4\frac{p}{q}$), we derive the
dimensions of the {\it h. w.} representations of $V_{3}^{(1,1)}$.

The main purpose of our discussion about the quantization of the classical
($PB^{\prime}$s) nonlocal $V_{3}^{(1,1)}$-algebra (\ref{13}) is to point out the
{\it differences} with the quantization of the $W_{3}$- and
$W_{3}^{(1,1)}$-algebras, and the {\it similarities} with the $PF$-algebra. The origin
of all these complications is the renormalization of the spins of the nonlocal
currents $V^{\pm}$, $s^{\pm}_{q}=s^{\pm}_{cl}-\frac{3}{2L}$, which makes the
singularities of the $V_{3}^{(1,1)}-OPE^{\prime}$s (\ref{35}) $L$-dependent. For
certain values of $L$, this requires to introduce new currents $V^{\pm}_{l}$ and
$W^{\pm}_{p}$ (see ref \cite{GSZ1}), in order to close the $OPE$-algebra. The
typical $PF$-feature is the replacing of the Lie commutators, with
an infinite sum of bilinears of generators, as in eqns. (\ref{36}) and
(\ref{37}). One might wonder whether the $V_{n+1}^{(1,1)}$-algebras (defined in
ref \cite{GSZ1}), exhibit similar features. Our preliminary results show that
the renormalization of the spins of the nonlocal currents $V^{\pm}_{(n+1)}$ is a
commun property of all $V_{n+1}^{(1,1)\prime}$s

\begin{center}
$s^{\pm}_{n}(q)=\frac{n+1}{2}\left( 1-\frac{1}{2k+n+1}\right)$.
\end{center}

\noindent
As usual, the spins of the local currents $W_{l+1}$ remain unchanged. All this
indicates that quantum $V_{n+1}^{(1,1)}$-algebras share many properties of
$V_{3}^{(1,1)}$. The construction of the {\it h. w.} representations of these
algebras, as well as the quantization of the $V$- and $WV$-algebras of other
types, say as in (\ref{20}) and in (\ref{21}), is an interesting open problem.
The same is valid for the simplest $U_{4}^{(1,2)}$-algebra, for the $UV$-algebra
$V_{4}^{(1,2)}$, of ref \cite{GSZ1}, and for the various explicit examples of
$U$- and $UV$-algebras, given in ref \cite{GSSZ}.

It is important to note, in conclusion, that the classification of the
{\it classical extensions} of the Virasoro algebra, described in this paper,
{\it does not solve} the problem of the classification of universality classes
in two dimensions. The complete solution of this challenge problem requires the
construction of the {\it h. w.} representations of the corresponding quantum
$W$-, $U$-, $V$ (and $WV$-, $UV$-)-algebras. We consider the above discussed
quantization of the $V_{3}^{(1,1)}$-algebras as a demonstration that
{\it relatively simple tools}, for the realization of this program,
{\it do exist}.

{\bf Acknowledgments}F. E. Mendon\c{c}a da Silveira thanks FAPESP, for financial support. G. M.
Sotkov thanks FAPESP, IFT-UNESP and DCP-CBPF, for financial support and
hospitalily. This work has been partially supported by CNPq.


\begin{thebibliography}{99}
\bibitem{BPZ}
A. Belavin, A. Polyakov and A. B. Zamolodchikov,
Nucl. Phys. {\bf B241} (1984) 333;
\bibitem{ABF}
G. E. Andrews, R. J. Baxter and P. J. Forrester,
J. Stat. Phys. {\bf 35} (1984) 193;
\bibitem{Kac1}
V. G. Kac,
Lecture Notes in Physics {\bf 94} (1979) 441;
\bibitem{Dots}
Vl. S. Dotsenko,
Nucl. Phys. {\bf B235} (1984) 54;
\bibitem{Huse}
D. A. Huse,
Phys. Rev. {\bf B30} (1984) 3908;
\bibitem{FQS1}
D. Friedan, Z. Qiu and S. Shenker,
Phys. Rev. Lett. {\bf 52} (1984) 1575;
Comm. Math. Phys. {\bf 107} (1986) 535;
\bibitem{FQS2}
D. Friedan, Z. Qiu and S. Shenker,
Phys. Lett. {\bf B151} (1985) 37;
\bibitem{KBT}
M. A. Bershadsky, V. G. Knizhnik and M. G. Teitelman,
Phys. Lett. {\bf B151} (1985) 31;
\bibitem{NS}
A. Neveu and J. H. Schwarz,
Nucl. Phys. {\bf B31} (1971) 86;
\bibitem{KMQ}
D. Kastor, E. Martinec and Z. Qiu,
Phys. Lett. {\bf B200} (1988) 434;
\bibitem{Rav}
F. Ravanini,
Mod. Phys. Lett. {\bf A3} (1988) 397;
\bibitem{ZF1}
A. B. Zamolodchikov and V. A. Fateev,
Sov. Phys. JETP {\bf 62} (1985) 215;
\bibitem{ZF2}
A. B. Zamolodchikov and V. A. Fateev,
Sov. Phys. JETP {\bf 63} (1985) 913;
\bibitem{ZF3}
A. B. Zamolodchikov and V. A. Fateev,
Nucl. Phys. {\bf B280} (1987) 644;
\bibitem{Zam}
A. B. Zamolodchikov,
Theor. Math. Phys. {\bf 65} (1986) 1205;
\bibitem{Kac-Moody}
V. Kac,
Funct. Anal. Appl. {\bf 1} (1967) 328;
R. V. Moody,
Bull. Amer. Math. Soc. {\bf 73} (1967) 217;
\bibitem{KZ}
V. G. Knizhnik and A. B. Zamolodchikov,
Nucl. Phys. {\bf B247} (1984) 83;
\bibitem{GW}
D. Gepner and E. Witten,
Nucl. Phys. {\bf B278} (1986) 493;
\bibitem{Ad}
M. Ademollo et all,
Phys. Lett. {\bf B62} (1976) 105;
Nucl. Phys. {\bf B111} (1976) 77;
\bibitem{BFK}
W. Bucher, D. Friedan and A. Kent,
Phys. Lett. {\bf B172} (1986) 316;
\bibitem{DiVec1}
P. Di Vecchia, J. L. Peterson and M. Yu,
Phys. Lett. {\bf B172} (1986) 211;
\bibitem{DiVec2}
P. Di Vecchia, V. G. Knizhnik, J. L. Peterson and P. Rossi,
Nucl. Phys. {\bf B253} (1985) 701;
\bibitem{KT}
V. G. Kac and I. Todorov,
Comm. Math. Phys. {\bf 102} (1985) 175;
\bibitem{Gepn}
D. Gepner,
Nucl. Phys. {\bf B290} (1987) 10;
\bibitem{FL}
V. A. Fateev and S. L. Lukyanov,
Int. J. Mod. Phys. {\bf A3} (1988) 507;
\bibitem{BG}
A. Bilal and J. L.  Gervais,
Nucl. Phys. {\bf B314} (1989) 646;
\bibitem{BBSS}
F. A. Bais, P. Bouwkneght, K. Schoutens and M. Surridge,
Nucl. Phys. {\bf B304} (1988) 371;
\bibitem{Bersh}
M. Bershadsky,
Comm. Math. Phys. {\bf 139} (1991) 71;
\bibitem{Poly1}
A. Polyakov,
Int. J. Mod. Phys. {\bf A5} (1990) 833;
\bibitem{O'Rai1}
J. Balog, L. Feh\'er, L. O$^{\prime}$Raifeartaigh, P. Forg\'acs and A. Wipf,
Ann. of Phys. {\bf 203} (1990) 76;
\bibitem{Bil1}
A. Bilal,
Nucl. Phys. {\bf B422} (1994) 258;
\bibitem{Bil2}
A. Bilal,
Comm. Math. Phys. {\bf 170} (1995) 117;
\bibitem{GSZ1}
J. F. Gomes, G. M. Sotkov and A. H. Zimerman,
Phys. Lett. {\bf 435B} (1998) 49;
\bibitem{GSZ2}
J. F. Gomes, G. M. Sotkov and A. H. Zimerman,
``Parafermionic Reductions of $WZW$ Model'', hep-th/9803234;
\bibitem{O'Rai2}
L. Feh\'er, L. O$^{\prime}$Raifeartaigh, P. Ruelle, I. Tsutsui and A. Wipf,
Phys. Rep. {\bf 222 N$^{o}_{\cdot}$1} (1992) 1;
\bibitem{BO}
M. Bershadsky and H. Ooguri,
Comm. Math. Phys. {\bf 126} (1989) 49;
\bibitem{GSSZ}
J. F. Gomes, F. E. Mendon\c{c}a da Silveira, G. M. Sotkov and A. H. Zimerman,
``Symmetries of Non-Abelian Toda Models'', in preparation;
\bibitem{Kac2}
V. G. Kac,
Funct. Anal. Appl. {\bf 3} (1969) 252;
\bibitem{LS}
A. N. Leznov and M. V. Saveliev,
Comm. Math. Phys. {\bf 74} (1980) 111;
Comm. Math. Phys. {\bf 89} (1983) 59.
\end{thebibliography}
\end{document}